\documentclass[showpacs,preprintnumbers,amssymb,nofootinbib]{revtex4}

\usepackage{graphicx}
\usepackage{bm} 
\def\be{\begin{equation}}
\def\ee{\end{equation}}
\def\beq{\begin{eqnarray}}
\def\eeq{\end{eqnarray}}

\def\bes{\begin{eqnarray}}
\def\ees{\end{eqnarray}}

\begin{document}

\preprint{NSF-KITP-06-04}

\title{Superradiant instability of large radius doubly spinning black rings}

\author{\'Oscar J. C. Dias\footnote{Also at KITP - Kavli
Institute for Theoretical Physics, UCSB, Santa Barbara, CA 93106,
USA.}} \email{odias@perimeterinstitute.ca}
 \affiliation{Perimeter Institute for Theoretical Physics,  \
 Waterloo, Ontario N2L 2Y5, Canada \\
  \\
Department of Physics, University of Waterloo,
 Waterloo, ON N2L 3G1, Canada }

\begin{abstract}
We point out that 5D  large radius  doubly spinning black rings
with rotation along $S^1$ and $S^2$ are afflicted by a robust
instability. It is triggered by superradiant bound state modes.
The Kaluza-Klein momentum of the mode along the ring is
responsible for the bound state. This is shown by analyzing the
boosted Kerr black string. The present kind of instability in
black strings and branes was first suggested by Marolf and Palmer
and studied in detail by Cardoso, Lemos and Yoshida.  We find the
frequency spectrum and timescale of this instability in the black
ring background,  and show that it is active for infinite radius
rings with large rotation along $S^2$. We identify the endpoint of
the instability and argue that it provides a dynamical mechanism
that introduces an upper bound in the rotation of the black ring.
To estimate the upper bound, we use the recent black ring model of
Hovdebo and Myers, with a minor extension to accommodate an extra
small angular momentum. This dynamical bound can be smaller than
the Kerr-like bound imposed by regularity at the horizon.
Recently, the existence of higher dimensional black rings is being
conjectured. They will be stable against this mechanism.
\end{abstract}


\maketitle

\section{Introduction}

It is not an easy problem to find exact black hole solutions with
a clear physical interpretation in gravity theories, but it is not
less difficult to prove or discard their uniqueness and stability.
In four dimensions, years of research ended with the conclusion
that the Kerr  black hole solution is stable and satisfies a
uniqueness theorem.

The higher dimensional arena, both in the vacuum Einstein theory
and in its extension to include the supergravity fields of low
energy string theory, has a broaden variety of black hole
solutions. For example, five-dimensional (5D) Einstein theory
allows not only the existence of black holes with
topology $S^3$ -- the Tangherlini black hole \cite{tang} and
the Myers-Perry black hole \cite{myersperry} --
but also the Emparan-Reall black ring  with topology $S^1\times
S^2$ \cite{EmpReall}, and extended objects known as black strings
\cite{HorStrom}.

This higher variety of solutions suggests that the uniqueness
theorem might not be valid in higher dimensions. The uniqueness of
the 5D Myers-Perry solution for black holes with
topology $S^3$,  and with two commuting spacelike Killing vectors
and a stationary timelike Killing vector
has been proven \cite{MorIda}. However, when other topologies and
Killing symmetries are considered, the
uniqueness theorem is violated as the discovery of
the black ring explicitly proves.  In five dimensions, a black
hole can rotate along two distinct planes. The rotating black ring
of \cite{EmpReall} has angular momentum only along $S^1$. Setting
one of the angular momenta equal to zero in the Myers-Perry black
hole, there is an upper bound for the ratio between the angular
momentum $J$ and the mass $M$ of the black hole, $J^2/M^3\leq
32/(27\pi)$, while for the black ring there is a lower bound in
the above ratio, $J^2/M^3\geq 1/\pi$, i.e., there is a minimum
rotation that is required in order to prevent the black ring from
collapsing. Now, what is quite remarkable is that for $1/\pi\leq
J^2/M^3\leq 32/(27\pi)$ there are Myers-Perry black holes and
Emparan-Reall black rings with the same values of $M$ and $J$.
Moreover, for the same values of the conserved charges, there are
actually two branches of black rings differing on the size of
their radius. The uniqueness in higher dimensions is further
violated when extra conserved charges and dipole charges are
included \cite{BRcharge,ElvEmp,BRDipole,BRnonBPS}. For our
purposes, it will be important to note that one of the parameters
that characterizes the Emparan-Reall black ring is  its radius,
$R$ \cite{EmpReall}. When this radius goes to zero the black ring
reduces to the Myers-Perry black hole, while in the
infinite-radius limit it yields a boosted black string. Recently,
a vacuum  black ring solution with angular momentum along $S^2$
(but no rotation along $S^1$) was found by Mishima and Iguchi
\cite{MishIguchi}. This solution was rediscovered and properly
interpreted by Figueras \cite{Figueras}. Since this
solution does not rotate along $S^1$, it is balanced by the
pressure of a conical disk. When its radius goes to zero the black
ring reduces to the Myers-Perry black hole, while in the
infinite-radius limit it yields a Kerr black string
\cite{Figueras}. The most general element of the vacuum black ring
family of solutions -- hereafter dubbed doubly spinning black ring
-- with rotation both along  $S^1$ and $S^2$ is still not yet
known. However, we know that it will reduce to  \cite{EmpReall} or
\cite{Figueras} when one of the rotations vanishes. Moreover, the
zero-radius limit of this doubly spinning black ring will be a 5D
Myers-Perry black hole with two angular momenta, and the
infinite-radius limit will yield a boosted Kerr black string.

Black hole solutions in higher dimensions bring also new
challenges to the stability issue. Some progress is being slowly
achieved. The stability of Tangherlini black holes
is established \cite{IshiKod}. However, in the higher dimensional arena it was
been progressively realized that instability is more the rule
rather than the exception. There are indeed several instabilities
that afflict  higher dimensional objects, and some of them can be
grouped in the following classes:

\noindent a) The Gregory-Laflamme instability \cite{grelaf}: This
gravitational instability is active in extended black objects like
black strings, black branes, and at least large radius black rings
\cite{EmpReaGR,boost}. The unstable modes are long wavelength
modes along the string, so this instability can be eliminated by
reducing the radius of the compactification along the string. For
a fixed radius, it is also suppressed by boosting the solution
along the string direction \cite{boost}.

\noindent b) Superradiant instability: Superradiant scattering,
where an incident wave packet can be reflected with a stronger
amplitude, can lead to an instability if, e.g., we have a
reflecting wall surrounding the black hole that scatters the
returning wave back toward the horizon.  In such a situation, the
wave will bounce back and forth, between the mirror and the black
hole, amplifying itself each time. The total extracted energy
grows exponentially until finally the radiation pressure destroys
the mirror.  This is Press and Teukolsky's black hole bomb, first
proposed in \cite{press} and studied in detail in \cite{bhb}. This
instability can arise with a natural `mirror' in a variety of
situations: a massive scalar field in a Kerr  background creates a
potential that can cause flux to scatter back toward the horizon
\cite{detweiler}; infinity in asymptotically AdS spaces also
provides a natural wall that leads, for certain conditions, to an
instability \cite{CardDiasAdS}; a wave propagating around spinning
black strings may similarly find itself trapped, because the
Kaluza-Klein momentum along the string can provide a potential
barrier at infinity
\cite{donny,cardosobraneinst,cardosobraneinst2}. The unstable
modes can be scalar, electromagnetic or gravitational. In general,
but with exceptions, this instability can be present for arbitrary
values of angular momentum of the geometry.

\noindent  c) Gyration instability: In the context of D1-D5-P
black strings, it was found that a spinning black string whose
angular momentum exceeds a certain critical value, decays into a
gyrating black string in which part of the original angular
momentum is carried by gyrations of the string  \cite{donny}. The
gyrating string has a helical profile traveling along the string
with the velocity of light. This instability is active for long
strings with large angular momentum.

\noindent  d) Ultra-spin instability: In six or higher spacetime
dimensions, there is no upper bound for the rotation of the
Myers-Perry black hole \cite{myersperry}. However, it was argued
\cite{ultra} that a Gregory-Laflamme-like instability will arise
to dynamically enforce a Kerr-like bound in these cases. While
this analysis does not directly apply in five dimensions, entropy
arguments suggest an analogous instability still exists and will
lead to the formation of a black ring if the angular momentum is
too large \cite{EmpReall}.

\noindent  e) Ergoregion instability: Geometries that have an
ergoregion but are horizon-free develop this kind of instability
\cite{friedman}. It occurs in rotating stars with an ergoregion
\cite{cominsschutz}, as well as in smooth non-supersymmetric
D1-D5-P geometries \cite{ergoross}.

In this paper we point out that 5D doubly spinning back rings can
be expected to suffer from the superradiant instability described
in item b), and we find its properties. This instability is
triggered by two factors that occur simultaneously. On one hand,
the KK momentum of the waves along the ring works effectively as a
massive term that provides a potential barrier at infinity. This
potential barrier allows the appearance of a potential well where
bound states get trapped. On the other hand, the bound state modes
can suffer superradiant amplification in the ergoregion. The
unstable modes are then waves that bounce back and forth between
the ergoregion and the potential barrier and that are amplified in
each scattering.  This kind of instability, where the bound states
are induced by the KK momentum, has first suggested by Marolf and
Palmer \cite{donny}, and explicitly studied by Cardoso, Lemos and
Yoshida \cite{cardosobraneinst,cardosobraneinst2}. This is the
same instability, but with a different source for the bound
states, as the one studied in
\cite{press,bhb,detweiler,CardDiasAdS}.

To fully study analytically this instability in the black ring
geometry we would have to separate the wave equation in this
background. However, it is not possible to do this separation even
in the scalar case and even in the black ring with rotation only
along $S^1$. This is explicitly shown in \cite{CardDiasYosh},
where the term that impedes this wave separation is identified.
This obstacle is closely connected with the fact that the
Hamiltonian-Jacobi equation for the geodesics is not also
separable \cite{EmpReallUnp}, and ultimately with the fact that
the solution does not have a Killing tensor \cite{EmpHerdUnp}. An
investigation of waves in this geometry will have to be done using
full 3D numerical simulations. Given this handicap (and the fact
that the finite radius doubly spinning black ring metric is still
not yet known), we do what can be done analytically: we consider
the large radius limit of the doubly spinning black ring -- which
yields a boosted Kerr black string -- and we verify that this
instability is active in this background. We then compactify the
string direction to find conclusions about the instability in
finite radius black rings.

Whenever one has an instability, we naturally ask what is its
endpoint. Usually, it is remarkably difficult to give a definite
answer to this question. For example, it still remains an open
issue the ultimate fate of the Gregory-Laflamme instability
\cite{GLend,Sorkin:2004qq}, of the gyration instability
\cite{donny}, and of the ergoregion instability \cite{ergoross}.
However, the superradiant instabilities are special, in the sense
that we usually can make strong statements  about their endpoint.
This follows from the fact that the growth of this type of
instability is fed by the rotational energy of the background
geometry which decreases during the process. In our case, we will
conclude that the boosted Kerr black string will release all its
angular momentum, and will settle down into a non-rotating boosted
black string. However, as a consequence of the KK momentum
quantization that occurs when we compactify a black string,  we
will argue that finite radius black rings will not loose all their
angular momentum along $S^2$. Therefore, in practice this means
that the superradiant instability provides a dynamical mechanism
that imposes an upper bound for the rotation of the black rings
along $S^2$. Note that due the the presence of other instabilities
on the system, like, e.g., the Gregory-Laflamme one, we expect
that a classical evolution of the black ring will proceed with the
contribution of all the instabilities that afflict the system. It
is not clear yet what is the exact endstate of this evolution.

The plan of the paper is organized as follows. In sections
\ref{sec:geometries}-\ref{sec:quantitative}  it is argued that 5D
doubly spinning black rings are expected to be unstable. In
section \ref{sec:geometries} we review the relevant black ring
geometries and their large radius limit. In section
 \ref{sec:Weq black string} we separate the Klein-Gordon equation in the
infinite radius black ring background, and we discuss the relevant
ingredients and properties of the instability. The quantitative
features of the instability, namely its frequency spectrum and its
timescale are computed in section \ref{sec:quantitative}, using
the matched asymptotic method. The stability analysis of the
higher dimensional boosted Myers-Perry black string is carried in
section \ref{sec:Hi dim}, where it is shown that they are stable
against the superradiant mechanism. The eventual implications of
this result to the stability of higher dimensional rings is then
discussed. Finally, in section \ref{discussion} we discuss the
results, with an emphasis given to the issue of the instability
endpoint. In particular, we argue that the instability discussed
in this paper provides a mechanism that effectively bounds the
rotation of doubly spinning black rings. We estimate the upper
bound, both on the rotation along $S^1$ and $S^2$, using the
recent black ring model of \cite{boost}, with a minor extension to
accommodate an extra small angular momentum.

\section{\label{sec:geometries}Black ring geometries and their large radius limit}

\subsection{The black ring with rotation along $\bm S^1$}

The first element of the family of 5D black rings  has found by
Emparan and Reall \cite{EmpReall}  and is described by (here we
write the metric in the form displayed in \cite{ElvEmp} after
using the factorization of \cite{EmpComp})
\begin{eqnarray}
 ds^2 =-\frac{F(y)}{F(x)}\left( dt+R \sqrt{\lambda
(\lambda-\nu)\frac{1+\lambda}{1-\lambda}}\,\,\frac{1+y}{F(y)}
d\psi \right)^2 + \frac{R^2}{(x-y)^2} F(x) {\biggl [}
-\frac{G(y)}{F(y)}d\psi
^2-\frac{dy^2}{G(y)}+\frac{dx^2}{G(x)}+\frac{G(x)}{F(x)}d\phi
^2{\biggl ]}, \label{metric S1}
\end{eqnarray}
where
\begin{eqnarray}
F(\xi) = 1+\lambda \xi\,,\qquad  G(\xi) = (1-\xi ^2)(1+\nu \xi)\,.
\end{eqnarray}
The coordinates $x$ and $y$ vary within the ranges,
\begin{eqnarray}
-1\leq x \leq 1\,, \qquad -\infty < y \leq -1\,. \label{range xy}
\end{eqnarray}
 To avoid conical singularities the period of the angular
coordinates $\phi$ and $\psi$ must be
\begin{eqnarray}
 \Delta \phi=\Delta
\psi=\frac{2\pi \sqrt{1-\lambda}}{1-\nu}\,, \label{period S1}
 \end{eqnarray}
and the parameters $\lambda$ and $\nu$ must satisfy the relation
\begin{eqnarray}
 \lambda=\frac{2\nu}{1+\nu ^2}\,, \qquad  0<\nu <1\,.
\label{equil cond S1}
\end{eqnarray}
  This condition guarantees that the rotation
of the ring balances the gravitational self-attraction of the
ring.

The black ring has a curvature singularity at $y= -\infty$. The
regular event horizon of topology $S^1\times S^2$ is at $y=-1/\nu$.
The ergosphere is located at $y=-1/\lambda$. The solution is
asymptotically flat with the spatial infinity being located at
$x=-1$ and $y=-1$. When $\lambda=\nu$ the geometry
 (\ref{metric S1}) describes a static black ring without rotation,
 whose dynamical equilibrium is ensured by a conical singular disk
whose pressure balances the self-gravitational attraction of the
ring.

We will be particularly interested in the large radius limit of
(\ref{metric S1}), which is defined by taking \cite{ElvEmp}
\begin{eqnarray}
 R \rightarrow \infty\,,\quad \lambda\rightarrow 0\,,\quad {\rm
and} \quad\nu \rightarrow 0\,, \label{limit S1}
 \end{eqnarray}
in the solution (\ref{metric S1}), and keeping $R\nu$ and
$\lambda/\nu$ finite. More particularly, take
\begin{eqnarray}
 R\nu=2M\,,\quad \frac{\lambda}{\nu}=\cosh ^2
\sigma\,, \quad r=-\frac{R}{y}\,,\quad \cos \theta=x\,,\quad
z=R\psi\,.\label{coord transf S1}
 \end{eqnarray}
 where $M$ and $\sigma$ are constants.
Then the black ring solution (\ref{metric S1}) goes over to the
boosted black string solution with horizon located at $r=2M$ and
with boost angle $\tanh\sigma=1/\sqrt{2}$. This particular value
of the boost angle is the one that follows from the balance
condition (\ref{equil cond S1}). This solution can also be
constructed by applying a Lorentz boost with angle $\sigma$ to the
geometry 4-dimensional (4D) Schwarzschild $\times \mathbb{R}$
\cite{BRcharge}. The parameter $M$ is the mass density of the
black string extended along the $z$ direction.

The black ring solution can be extended in order to include
electric charge \cite{BRcharge,ElvEmp}, as well as magnetic dipole
charge \cite{BRDipole}.
 The most general known seven-parameter family of non-supersymmetric
black ring solutions, that includes the above solutions as special
cases, was presented in \cite{BRnonBPS}. It is characterized by
three conserved charges, three dipole charges, two unequal angular
momenta, and a parameter that measures the deviation from the
supersymmetric configuration. In the black ring background, the
Penrose process was discussed in \cite{NozMaeda},  perturbations
analysis in the large radius limit was carried in
\cite{CardDiasYosh}, and a ultrarelativistic boost of the black
ring was considered in \cite{ortaggio}. The algebraic classification of
the black ring is done in
\cite{algebraic}. There are also
supersymmetric black rings that we will not discuss in this paper.
A recent detailed discussion of the supersymmetric black ring
system can be found, e.g., in \cite{ElvEmpMatReallGaunt}.

\subsection{\label{brS2}The black ring with rotation along $\bm S^2$}

The 5D rotating black ring with rotation along $S^2$ (and no
rotation along $S^1$) was found by Mishima and Iguchi
\cite{MishIguchi} and Figueras \cite{Figueras}. In its most
appropriate form for physical interpretation \cite{Figueras},
the line element is
 \begin{eqnarray}
 ds^2&=&-\frac{H(\lambda,y,x)}{H(\lambda,x,y)}\left
(dt-\frac{\lambda\, a\, y (1-x^2)}{H(\lambda,y,x)}\,
d\phi \right )^2 \nonumber \\
& & + \frac{R^2 \,H(\lambda,x,y) }{(x-y)^2} {\biggl [}
-\frac{dy^2}{(1-y^2)F(\lambda,y)} -
\frac{(1-y^2)F(\lambda,x)}{H(\lambda,x,y)}\, d\psi^2 +
\frac{dx^2}{(1-x^2)F(\lambda,x)} +
\frac{(1-x^2)F(\lambda,y)}{H(\lambda,y,x)}\, d\phi^2 {\biggl ]},
\label{metric S2}
 \end{eqnarray}
 where
\begin{eqnarray}
F(\lambda,\xi) = 1+\lambda \xi + \left(
\frac{a\xi}{R}\right)^2\,,\qquad  H(\lambda,\xi_1,\xi_2) = 1+
\lambda \xi_1 + \left( \frac{a\xi_1 \xi_2}{R}\right)^2\,.
\end{eqnarray}
The coordinates $x$ and $y$ vary in the intervals defined in
(\ref{range xy}). To avoid conical singularities at $x=-1$ and
$y=-1$, the period of the angular coordinates $\phi$ and $\psi$
must be given by
\begin{eqnarray}
 \Delta \phi=\Delta
\psi=\frac{2\pi}{ \sqrt{1-\lambda+a^2/R^2}}\,. \label{period S2}
 \end{eqnarray}
However, the solution has a conical singularity at $x=1$ that
signals the presence of a conical singular disk that balances the
self-gravitational attraction of the ring. The parameter $\lambda$
and $a$ must satisfy the relation
\begin{eqnarray}
 \frac{2a}{R}<\lambda <1+ \frac{a^2}{R^2}\,,
\label{equil cond S2}
\end{eqnarray}
where the lower bound guarantees that there is a horizon and the
upper bound ensures the absence of closed timelike curves. This
black ring has a curvature singularity, and a regular event
horizon with topology $S^1\times S^2$. It is asymptotically flat
with the spatial infinity being located at $x=-1$ and $y=-1$.
Moreover, the black ring has rotation along the azimuthal
direction $\phi$ of the $S^2$, and it also has an ergoregion.

Again, we will be interested in the large radius limit of
(\ref{metric S2}), which is defined by taking \cite{ElvEmp}
\begin{eqnarray}
 R \rightarrow \infty\,,\quad {\rm
and} \quad \lambda\rightarrow 0\,, \label{limit S2}
 \end{eqnarray}
in the solution (\ref{metric S2}), and keeping $R\lambda$ fixed.
More concretely, take
\begin{eqnarray}
 R\lambda=2M\,,\quad r=-\frac{R}{y}\,,\quad \cos \theta=x\,,\quad
z=R\psi\,, \label{coord transf S2}
 \end{eqnarray}
 where $M$ is a constant.
Then the black ring solution (\ref{metric S2}) goes over to the
Kerr black string solution extended along the $z$ direction, with
mass density $M$ and rotation parameter $a$ along the azimuthal
$\phi$ direction. This solution can also be constructed by adding
a flat direction to the  4D  Kerr black hole, yielding  4D  Kerr
$\times \mathbb{R}$.

\subsection{The large radius doubly spinning black ring: boosted spinning black string}

The most general 5D black ring solution -- the doubly spinning
black ring -- is not yet known. This more general element of the
family will have rotation both along the plane of the ring (i.e.,
along the $S^1$ direction parameterized by $\psi$) that will
balance the self-gravitation of the ring, and along the azimuthal
$\phi$ direction of the $S^2$. Therefore, when the rotation along
$S^2$ vanishes this doubly spinning solution reduces to
(\ref{metric S1}), while when it is the rotation along $S^1$ that
is absent it reduces to (\ref{metric S2}).

\begin{figure}[h]
\begin{center}
 {\includegraphics{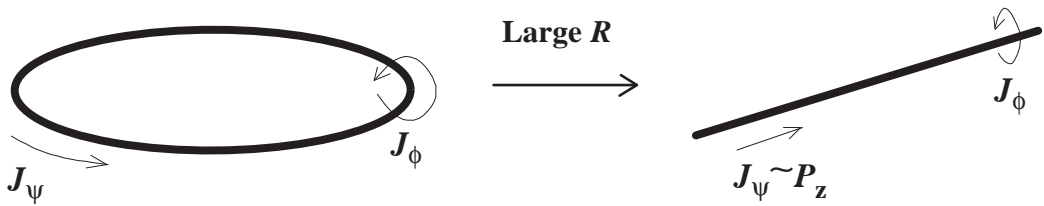}}
\end{center}
\caption{In the large radius limit, the doubly spinning black ring
yields a boosted Kerr black string.}
 \label{fig:br}
\end{figure}

Although we do not still know the line element of the doubly
spinning black ring, we do know the geometry that describes its
large radius limit. Indeed, it is not difficult to convince
ourselves (see Fig. \ref{fig:br}) that in this limit it will be
described by a boosted Kerr black string extended along the
$z$-direction. This black string is characterized by the mass
density $M$, by the boost angle $\sigma={\rm
arctanh}(1/\sqrt{2})$, and by the rotation parameter $a$ along the
azimuthal $\phi$-direction. This can also be constructed by adding
a flat direction $z$ to the  4D  Kerr black hole, and then
applying a Lorentz boost to it, $dt\rightarrow \cosh\sigma
dt+\sinh\sigma dz$ and $dz\rightarrow \sinh\sigma dt+\cosh\sigma
dz$. This yields the boosted Kerr black string geometry,
\begin{eqnarray}
ds^2&=&-\left( 1-\frac{2M r \cosh^2\sigma }{\Sigma}\right) dt^2+
\frac{2M r \sinh(2\sigma) }{\Sigma} \,dt dz + \left( 1+\frac{2M r
\sinh^2\sigma }{\Sigma}\right)dz^2 +\frac{\Sigma}{\Delta}dr^2
 +\Sigma d\theta^2  \nonumber \\
 & & +\frac{(r^2+a^2)^2-\Delta a^2
\sin^2\theta}{\Sigma}\,\sin^2\theta d\phi^2
  -\frac{4M r \cosh\sigma }{\Sigma}\,a\sin^2\theta dt d\phi
  -\frac{4M r \sinh\sigma }{\Sigma-2Mr}\,
  a\sin^2\theta dz d\phi\,,
   \label{metric}
\end{eqnarray}
 where
 \begin{eqnarray}
 \Delta &=& r^2+ a^2 -2Mr \,, \nonumber \\
 \Sigma &=& r^2+ a^2\cos^2\theta \,.
   \label{metric parameters}
\end{eqnarray}
 When we set $a=0$ and $\sigma={\rm
arctanh}(1/\sqrt{2})$, we get a boosted black string that
describes a large radius skinny  balanced black ring that is also
found by taking the limit (\ref{limit S1}), (\ref{coord transf
S1}) in (\ref{metric S1}). Unbalanced rings give boosted black
strings with a general $\sigma$. In the data examples that we will
give in the paper, we will fix $\sigma$ to have the balanced
value.
 On the other side, if we set $\sigma=0$ we
 get a Kerr black string that describes a
large radius skinny black ring that is also obtained by taking the
limit (\ref{limit S2}), (\ref{coord transf S2}) in
 (\ref{metric S2}).

The boosted Kerr black string has a curvature singularity at
$r=0$, a Cauchy horizon $r_-$ and an event horizon $r_+$ at
$r_{\pm}=M\pm\sqrt{M^2-a^2}$, and an ergosurface at
$r_{e}=M\cosh^2\sigma + \sqrt{M^2\cosh^4\sigma-a^2\cos^2\theta}$.
To avoid naked singularities, the  rotation is constrained to be
$a\leq M$.

Three important parameters of the boosted Kerr black string are
the angular velocity of the horizon along the $\phi$-coordinate,
$\Omega_\phi=-\frac{g_{t\phi}}{g_{\phi\phi}}{\bigl |}_{r=r_+}$,
the area per unit length of the horizon, $A_H$, and the
temperature of its horizon, $T_H$, given by
 \begin{eqnarray}
 & & \Omega_\phi=\frac{a\cosh\sigma}{r_+^2+a^2} \,,
 \nonumber \\
 & & A_H=4\pi (r_+^2+a^2) \cosh \sigma \,, \nonumber \\
& & T_H=\frac{r_+-r_-}{4\pi (r_+^2+a^2) \cosh \sigma}\,.
 \label{boostedbs TVA}
\end{eqnarray}
For completeness note that the linear velocity of the horizon along the string direction,
$V_z=-\frac{g_{tz}}{g_{zz}}{\bigl |}_{r=r_+}$, is given by $V_z=-\tanh \sigma$.

\section{\label{sec:Weq black string}The wave equation of the
boosted Kerr black string. Properties of the instability}

In this section we will show that a boosted Kerr black string is
unstable, against massless field perturbations, due to the
combined effect of the superradiant mechanism and of the presence
of an effective reflective wall sourced by the KK momentum of the
mode along the string direction. This wall provides the arena for
the existence of bound states. The mechanism at play in this
instability was recently described in
\cite{donny,cardosobraneinst,cardosobraneinst2,zelmisnerunruhbekenstein,bhb,CardDiasAdS}),
and is active in {\it some}  extended rotating black objects. The
massless scalar field acquires an effective mass (due to the KK
momentum in the extra dimension) and this makes the wave bounce
back and forth. In each scattering the wave suffers superradiant
amplification in the ergosphere and this leads to an instability
\cite{donny,cardosobraneinst}. This superradiant instability will
also be present in the case of electromagnetic and gravitational
waves.

We extend the analysis carried in
\cite{cardosobraneinst,cardosobraneinst2} for Kerr black strings
to the boosted case, and then we further extend this analysis to
the black ring case. Generically, the results for boosted black
strings are in a way the same as the ``boosted results'' of
unboosted black strings. However, we must be cautious since some
exceptions to this rule are already known. For example, a boosted
black string has an ergoregion and the Penrose process can occur
in this region, while in the unboosted case this phenomena is not
present \cite{NozMaeda}. As another example, it was recently shown
\cite{boost} that Sorkin's critical dimension \cite{Sorkin:2004qq}
-- below which stable black strings and small black holes on a
compact circle can coexist -- is boost dependent and actually
vanishes for large boosts. Therefore, the boosted results do not
seem to follow trivially from the unboosted ones. As explained
in \cite{boost}, this is intimately connected with the different
boundary conditions in the two systems.

\subsection{\label{sec:separation}Separation of the wave equation}

 The evolution of a scalar field $\Phi$ in the background of
(\ref{metric}) is governed by the curved space Klein-Gordon
equation, $\nabla_{\mu}\nabla^{\mu}\Phi=0$. It is appropriate to
use the separation ansatz
 \begin{eqnarray}
  \Phi= e^{-i \omega t}e^{i m \phi} e^{-i \kappa z}S^m _l(\theta)\Psi(r)
\,, \label{ansatz}
 \end{eqnarray}
where $S^m _l(\theta)$ are spheroidal angular functions, and the
azimuthal number $m$ takes on integer (positive or negative)
values. For our purposes, it is enough to consider positive
$\omega$'s in (\ref{ansatz}). Inserting this ansatz into the
Klein-Gordon equation, we get the following angular and radial
wave equations for $S^m _l(\theta)$ and $\Psi(r)$,
\begin{eqnarray}
& &\hspace{-1cm} \frac{1}{\sin \theta}\partial_{\theta}\left (
\sin \theta
\partial_{\theta} S^m _l \right )  + \left [  a^2 (\omega^2-\kappa^2)
\cos^2 \theta-\frac{m^2}{\sin ^2{\theta}}+\lambda_{lm} \right ]S^m
_l =0\,, \label{separation ang}
\\
& &\hspace{-1cm} \Delta  \partial_r \left ( \Delta \partial_r \Psi
\right ) - \Delta {\bigl [}  \kappa^2 r^2 +a^2\omega^2-2\omega m a
\cosh\sigma +\lambda_{lm}  {\bigr ]} \Psi + {\biggl [}  {\bigl
[}\omega(r^2+a^2)-m a \cosh\sigma {\bigr ]}^2
\nonumber \\
& &  \hspace{0.5cm}
 + 2M r(r^2+a^2)\cosh^2\sigma
 {\bigl [} \omega-\kappa\tanh\sigma {\bigr ]}^2  -2Mr(r^2+a^2)\omega^2
 -m^2a^2 \sinh^2\sigma +4\kappa m a M r \sinh\sigma {\biggr ]}\Psi
=0\,,
 \label{separation rad}
\end{eqnarray}
where $\lambda_{lm}$ is the separation constant that allows the
split of the wave equation, and is found as an eigenvalue of
(\ref{separation ang}). For small $a^2(\omega^2-\kappa^2)$, the
regime we shall be interested on, the eigenvalues associated with
the spheroidal wavefunctions $S^m _l$ are \cite{Teukolsky}
\begin{eqnarray}
\lambda_{lm}=l(l+1)+{\cal O}{\bigl ( }a^2(\omega^2-\kappa^2){\bigr
)} \approx l(l+1)\,,
 \label{eigenvalues}
\end{eqnarray}
where the integer $l$ is constrained to be $l\geq |m|$.

\subsection{\label{sec:bound cond}Boundary conditions. Necessary conditions for the instability.}

Together with the wave equation, we must also specify the
appropriate boundary  conditions for the instability problem. The
features of these boundary conditions reveal the two ingredients
responsible for the presence of the instability. We are interested
in wave perturbations that develop in the vicinity of the horizon
and that propagate both into the horizon and out to infinity. Our
boundary conditions therefore require only ingoing flux at the
horizon and only outgoing waves at infinity.

At the horizon, the second set of terms proportional to $\Delta$
in (\ref{separation rad}) can be neglected and the radial wave
equation reduces to
\begin{eqnarray}
& & \Delta  \partial_r \left ( \Delta \partial_r \Psi \right ) +
(r_+^2+a^2)^2 \cosh^2\sigma \left(\omega-\kappa\tanh \sigma-
\frac{m a}{(r_+^2+a^2)\cosh \sigma}\right)^2 \Psi =0\,, \qquad
{\rm as \,\,} r\rightarrow r_+\,,
 \label{rad horizon}
\end{eqnarray}
where we made use of $\Delta(r_+)=0$ which implies
$2Mr_+=(r_+^2+a^2)$. At infinity, the  radial wave equation
(\ref{separation rad}) is dominated by
\begin{eqnarray}
  \partial_r \left ( r^2 \partial_r \Psi \right )
+ (\omega^2-\kappa^2)r^2   \Psi =0\,, \qquad {\rm as \,\,}
r\rightarrow \infty\,,
 \label{rad infinity}
\end{eqnarray}
where we used $\Delta \sim r^2$ as $r \rightarrow \infty$. Here,
or in (\ref{separation rad}), we realize that the KK momentum
$\kappa$ contributes effectively as a mass term, $-\kappa^2 r^2$,
 in the wave equation.

The solutions of these two equations that satisfy our boundary
conditions are
\begin{equation}
\Psi(r)\sim
 \left\{
\begin{array}{ll}
(r-r_+)^{-i \varpi}=e^{-i \varpi \ln (r-r_+)}\,, \qquad  {\rm as}
\quad r\rightarrow r_+ \,, \\
r^{-1}\,e^{+i \sqrt{\omega^2-\kappa^2} \,r}\,, \qquad   {\rm as}
\quad r\rightarrow \infty \,,
\end{array}
\right.
\label{bound cond}
\end{equation}
where we have defined
\begin{eqnarray}
\varpi &\equiv& \frac{(r_+^2+a^2) \cosh\sigma}{r_+-r_-}
\left(\omega-\kappa\tanh \sigma-
\frac{m a}{(r_+^2+a^2)\cosh \sigma}\right) \nonumber \\
&\equiv& \frac{1}{4\pi T_H} \left( \omega
 -\omega_{\rm sup} \right)\,,
    \label{superrad factor}
\end{eqnarray}
with
\begin{eqnarray}
\omega_{\rm sup} = \kappa \tanh \sigma
 +\frac{m \Omega_{\phi} }{\cosh^2\sigma} \,,
    \label{wsup}
\end{eqnarray}
and $T_H$ and $\Omega_{\phi}$ are, respectively, the temperature
and the angular velocity of the horizon of the boosted Kerr black
string defined in (\ref{boostedbs TVA}).

The features of the boundary conditions (\ref{bound cond}) reveal
the two ingredients responsible for the presence of the
instability. Indeed, at infinity we identify the presence of the
KK term or mass term, $\kappa^2$, that is responsible for the
effective reflective wall. On the other side, at the horizon one
identifies the presence of the so-called superradiant factor
$\varpi$. When the frequency of the wave is such that  $\varpi$ is
negative,
\begin{eqnarray}
\omega< \omega_{\rm sup} \,.
 \label{super cond}
\end{eqnarray}
one is in the superradiant regime, and the amplitude of a
(co-rotating) wave is amplified after each scattering.

It is important to understand why it is $\Psi \sim e^{-i \varpi
\ln (r-r_+)}$ and not  $\Psi \sim e^{+ i \varpi \ln (r-r_+)}$ that
describes always an ingoing wave at the horizon, since this is the
key source of superradiance. It follows from (\ref{ansatz}) and
(\ref{bound cond}) that at the horizon the wave solution behaves
as $\Phi(t,r) {\bigl |}_{r\rightarrow r_+}\sim e^{-i\omega t}
e^{-i \varpi \ln (r-r_+)} $. The phase velocity of the wave is
then $v_{\rm ph} \propto -\frac{\omega}{\varpi}$. Now, the value
of this phase velocity can be positive or negative depending on
the value of $\omega$ (when we fix the other parameters), so one
might question if the first line of  (\ref{bound cond}) really
describes {\it always} an ingoing wave. As in other applications,
what is relevant to find the {\it physical} ingoing wave solution
at the horizon is the group velocity of the wave rather than its
phase velocity.  The normalized group velocity, $v_{\rm gr}$, at
the horizon is $v_{\rm gr}=4\pi T_H \, \frac{d(-\varpi)}{d \omega}
= -1$. This is a negative value that signals that the horizon wave
solution in (\ref{bound cond}) always represents an ingoing wave
{\it independently} of the value of $\omega$, and is thus the
correct physical boundary condition. However, note that in the
superradiant regime (\ref{super cond}), the phase velocity is
positive and so  waves appear as outgoing to an inertial observer
at spatial infinity -- energy is in fact being extracted.

Notice that (since we are working with positive $\omega$)
superradiance will occur only for positive $m$, i.e., for waves
that are co-rotating with the black hole. Indeed, the wave
function is given by $\Phi (t,\phi) \sim e^{ i\omega t+i m \phi}$.
The phase velocity along the angle $\phi$ is then $v_{\phi} =
\omega/m$, which for $\omega > 0$ and $m > 0$ is positive, i.e.,
is in the same sense as the angular velocity $\Omega_{\phi}$ of
the black string.

\subsection{\label{sec:schr 4D} The radial wave equation as a
Schr$\bm \ddot{\rm \bf o}$dinger equation. Necessary and sufficient
conditions for the instability.}

\begin{figure}[t]
\begin{center}
{\includegraphics{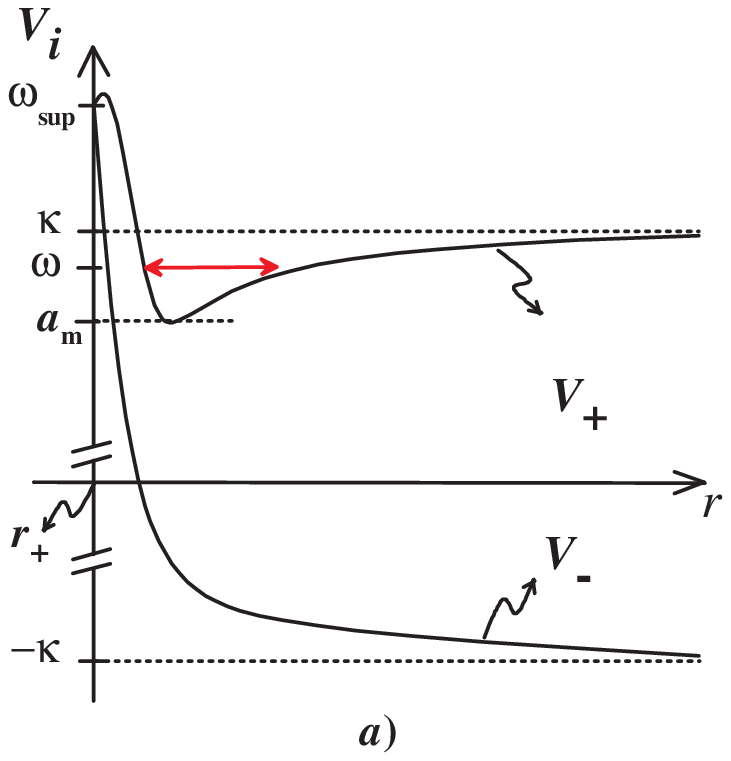}\qquad \qquad
\includegraphics{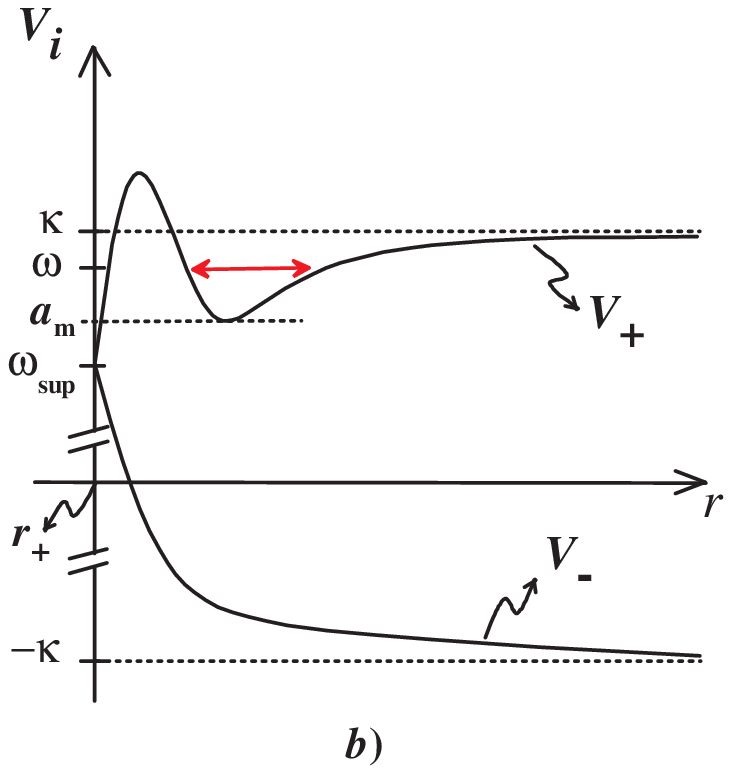}}
\end{center}
\caption{{\bf a)}Qualitative shape of the potentials $V_+$ and
$V_-$ for a boosted Kerr black string in which an instability is
present. An example of data that yields this kind of potentials is
$(2M=1\,,a=0.499999\,,\tanh\sigma=1/\sqrt{2}\,,\kappa=1.85\,,l=m=1)$.
Potentially unstable modes are those whose frequency satisfies
$a_{\rm m}<\omega< \kappa$. Thus, they are bound states of the
potential well in $V_+$ that satisfy the superradiant condition
$\omega< \omega_{\rm sup}$. \\
 {\bf b)} Qualitative shape of the potentials $V_+$ and $V_-$ for
a case in which the Kerr black string is stable. An example of
data that yields this kind of potentials is
$(2M=1\,,a=0.4\,,\tanh\sigma=1/\sqrt{2}\,,\kappa=1.85\,,l=m=1)$.
The bound state modes with $a_{\rm m}<\omega< \kappa$ are stable
because they do not satisfy the superradiant condition $\omega<
\omega_{\rm sup}$, where $\omega_{\rm sup}$ is defined in
(\ref{wsup}).
 }
 \label{fig:potential 4D}
\end{figure}

The necessary conditions -- superradiant regime and presence of
the KK effective mass term -- are not, in general, sufficient to
guarantee that a instability develops in the geometry as we shall
conclude in section \ref{sec:Hi dim}. Necessary and sufficient
conditions are that the superradiant regime is active and that
{\it bound states} are present. The requirement of these
conditions is better understood by re-writting the radial wave
equation (\ref{separation rad}) in the Schr$\ddot{\rm
o}$dinger-like form,
 \begin{eqnarray}
  \partial_{r_{\ast}}^2\chi-V\chi=0 \,,\qquad {\rm with} \:\:\:\:
  V=-\gamma (\omega-V_+) (\omega-V_-)
  \,,\qquad  \Psi=(r_+^2+a^2)^{-1/2}\chi \,,
  \label{Schr aux 4D}
 \end{eqnarray}
and $r_{\ast}$ is the usual tortoise coordinate $r_{\ast}$. The
explicit form of $\gamma >0$ and $V_{\pm}$ will be given in great
detail in (\ref{V+V- hiDIM}), where we will also consider the
higher dimensional case. At the moment, we are interested in a
plot of the behavior of the potentials $V_{\pm}$. The asymptotic
behavior of $V_{\pm}$ is
 \begin{eqnarray}
  \lim_{r\rightarrow r_+} V_{\pm}= \omega_{\rm sup}\ \,, \qquad
  \lim_{r\rightarrow \infty} V_{\pm}= \pm |\kappa| \,. \label{lim V}
 \end{eqnarray}
 There are only
two distinct cases that we sketch in Fig.
 \ref{fig:potential 4D}.(a) (unstable case) and
in Fig. \ref{fig:potential 4D}.(b) (stable case). In these
figures, when  $\omega$ is above $V_+$ or below $V_-$ (allowed
regions), the solutions have an oscillatory behavior. In those
intervals where $\omega$ is in between the curves of  $V_+$ and
$V_-$ (forbidden regions), the solutions have a real exponential
behavior. In both Figs. \ref{fig:potential 4D}.(a) and
\ref{fig:potential 4D}.(b), the potential $V_+$ has a well that is
limited at infinity by the KK momentum potential barrier of height
$\kappa$. Modes with a frequency greater than the bottom of the
well, $a_{\rm m}$, and smaller than $\kappa$, $a_{\rm m}<\omega<
\kappa$, are bound states.  If these bound state frequencies
further satisfy the superradiant condition (\ref{super cond}), an
instability will settle down: the waves will bounce back and forth
between the KK wall at infinity and the black hole, amplifying
itself each time (Fig. \ref{fig:potential 4D}.(a)). If the bound
state frequencies do not satisfy the superradiant condition
(\ref{super cond}) the modes will bounce back and forth in the
potential well, but each time part of the wave will tunnel to the
horizon until the full mode is completely absorbed by the horizon
(Fig. \ref{fig:potential 4D}.(b)).

Here we note that when we set the rotation  parameter to zero,
$a=0$ (boosted Schwarzschild black string), we always get
potentials of the form represented in Fig. \ref{fig:potential
4D}.(b),  for any $M\,,\sigma$ and $\kappa\,,l\,,m$. In particular
this means that the bound states of the boosted black string are
always damped modes since the  superradiant condition (\ref{super
cond}) is never satisfied. This was first noticed in
\cite{NozMaeda,CardDiasYosh}, where it was shown that superradiant
scattering is never possible in a boosted  Schwarzschild black
string, although the particle analogue of this phenomena -- the
Penrose process -- can occur. Ultimately, this is because one can
construct, with a combination of $\partial/\partial t$ and $\partial /\partial z$,
 a Killing vector field that is everywhere timelike outside the horizon  \cite{rossKV}.

The KK momentum is a necessary condition for the existence of
bound states, but is not a sufficient condition as we shall see in
section  \ref{sec:Hi dim}, where we will conclude that the
Myers-Perry black string is stable due to the lack of bound
states.

\section{\label{sec:quantitative} Frequency spectrum. Instability timescale.}

In the last section we concluded that some boosted Kerr black
strings are unstable to scalar field perturbations that might
develop in the vicinity of the horizon, and we identified the
relevant ingredients for this process. However, we do not still
know the quantitative features of the instability, namely we do
not know the allowed spectrum of real frequencies of the bound
states, and what is the instability timescale.

In this section we will address these issues. The method we shall
use here, known as matched asymptotic expansion, has been widely
used with success for the computation of scattering cross-section
of black holes \cite{staro1}, and also for computing instabilities
in the Kerr background \cite{detweiler,bhb,CardDiasAdS}. We will
assume that the Compton wavelength of the scalar particle is much
larger than the typical transverse size of the black string; we
divide the space outside the event horizon in two regions, namely,
a near-region and a far-region. These two regions have a
overlapping region where we can match their wave solutions to get
a solution to the problem. This allows the analysis of the
instability properties. When the correct boundary conditions are
impose, namely only ingoing flux at the horizon and only outgoing
waves at infinity, we get a defining equation for $\omega$. The
stability or instability of the spacetime depends basically on the
sign of the imaginary component of $\omega$.

\subsection{\label{sec:BH Near region}The near-region solution}
First, let us focus on the near-region in the vicinity of the
horizon, $r-r_+ \ll 1/\omega$. We work in the regime $\omega^2
r_+^2 \ll 1$, $\omega^2 a^2 \ll 1$, $\kappa^2 r_+^2 \ll 1$ and
$\sigma$ not too large (which is definitely the case for balanced
rings), and define the new variable
\begin{eqnarray}
y=\frac{r-r_+}{r-r_-}\,,
\end{eqnarray}
whose range is $0\leq y \leq 1$. The horizon is at $y=0$ and
infinity is at $y=1$.
 We have $\Delta=(r_+ - r_-)^2 y /(1-y)^2$, $\Delta
\partial_r=(r_+ - r_-) y\partial_y$.
In the near-region and in the regime we work, we can neglect the
contribution of the terms coming from $\kappa^2 r^2
+a^2\omega^2-2\omega m a \cosh\sigma$ in (\ref{separation rad}),
when comparing them with $\lambda_{ml}\simeq l(l+1)$. In this case
we can find an analytical near-horizon solution for the wave
equation. More concretely, in the near-region, the radial wave
equation (\ref{separation rad}) can then be written as
\begin{eqnarray}
& &  \hspace{-1cm} y(1-y)\partial ^2 _y \Psi +(1-y)
\partial _y \Psi  + \left [ -\frac{l(l+1)}{1-y}+
\frac{1-y}{y}\varpi^2\right ]\Psi =0\,,
 \label{near wave eq}
 \end{eqnarray}
where we have introduced the superradiant factor $\varpi$ as
defined in (\ref{superrad factor}).
 Through the definition
\begin{eqnarray}
\Psi=z^{i \,\varpi} (1-y)^{l+1}\,F \,,
 \label{hypergeometric function}
\end{eqnarray}
the near-region radial wave equation becomes
\begin{eqnarray}
 y(1-y)\partial_y^2 F+ {\biggl [} (1+i\, 2\varpi)-\left [
1+2(l+1)+ i\, 2\varpi \right ]\,y {\biggr ]}
\partial_y F - \left [ (l+1)^2+ i \,2\varpi (l+1)\right
]F=0\,.
 \label{near wave hypergeometric}
\end{eqnarray}
This wave equation is a standard hypergeometric equation
\cite{abramowitz} of the form
 \begin{eqnarray}
y(1-y)\partial_y^2 F+[c-(a+b+1)y]\partial_y F-a b F=0
\,,\end{eqnarray}
 with
\begin{eqnarray}
 a=l+1+i\,2\varpi \,,  \qquad b=l+1 \,, \qquad
c=1+ i\,2\varpi \,.
 \label{hypergeometric parameters}
\end{eqnarray}
The general solution of this equation in the neighborhood of $z=0$
is $A\, y^{1-c} F(a-c+1,b-c+1,2-c,y)+B\, F(a,b,c,y)$. Using
(\ref{hypergeometric function}), one finds that the most general
solution of the near-region equation is
\begin{eqnarray}
 \Psi = A\, y^{-i\,\varpi}(1-y)^{l+1}
F(a-c+1,b-c+1,2-c,y)+B\,y^{i\,\varpi}(1-y)^{l+1} F(a,b,c,y) \,.
 \label{hypergeometric solution}
\end{eqnarray}
The first term represents an ingoing wave at the horizon $y=0$,
while the second term represents an outgoing wave at the horizon
[recall the discussion after (\ref{super cond})]. We are working
at the classical level, so there can be no outgoing flux across
the horizon, and thus one sets $B=0$ in (\ref{hypergeometric
solution}). Note that one then gets the near-horizon solution
written in  (\ref{bound cond}). Later on, to do the matching
between the near and far regions, we will be interested in the
large $r$, $y\rightarrow 1$, behavior of the ingoing near-region
solution. To achieve this aim one uses the $y \rightarrow 1-y$
transformation law for the hypergeometric function
\cite{abramowitz},
\begin{eqnarray}
 F(a\!-\!c\!+\!1,b\!-\!c\!+\!1,2\!-\!c,y)&=&
(1\!-\!y)^{c-a-b}
\frac{\Gamma(2-c)\Gamma(a+b-c)}{\Gamma(a-c+1)\Gamma(b-c+1)}
 \,F(1\!-\!a,1\!-\!b,c\!-\!a\!-\!b\!+\!1,1\!-\!y) \nonumber \\
 & & + \frac{\Gamma(2-c)\Gamma(c-a-b)}{\Gamma(1-a)\Gamma(1-b)}
 \,F(a\!-\!c\!+\!1,b\!-\!c\!+\!1,-c\!+\!a\!+\!b\!+\!1,1\!-\!y),
 \label{transformation law}
\end{eqnarray}
and the property $F(a,b,c,0)=1$. The large $r$ behavior of the
ingoing wave solution in the near-region is then given by
\begin{eqnarray}
\Psi \sim A\,\Gamma(1-i\,2\varpi){\biggl [} \frac{(r_+
-r_-)^{-l}\,\Gamma(2l+1)}{\Gamma(l+1)\Gamma(l+1-i\,2\varpi)}\:
r^{l} +\frac{(r_+
-r_-)^{l+1}\,\Gamma(-2l-1)}{\Gamma(-l)\Gamma(-l-i\,2\varpi)}\:
r^{-l-1} {\biggr ]}.
 \label{near field large r}
\end{eqnarray}

\subsection{\label{sec:BH Far region}The far region solution}
In the far-region, $r-r_+ \gg r_+$, and for $\omega a \ll 1$, the
wave equation (\ref{separation rad}) reduces to
\begin{eqnarray}
\partial ^2 _r (r \Psi) +{\biggl [} \omega ^2-\kappa^2
+\frac{r_+ + r_-}{r}(\omega \sinh \sigma-\kappa \cosh \sigma)^2
 -\frac{l(l+1)}{r^2} {\biggr ]}(r \Psi) =0 \,.
 \label{far wave eq aux}
 \end{eqnarray}
If we define
\begin{eqnarray}
\eta ^2 & \equiv & \kappa^2-\omega ^2 \,,\nonumber \\
\rho & \equiv &
\frac{(r_+ + r_-) (\omega \sinh\sigma
-\kappa \cosh \sigma)^2} {2\eta} \,, \nonumber \\
\chi & = & 2 \eta r\,,
          \label{far wave parameters}
\end{eqnarray}
equation (\ref{far wave eq aux}) is written as
\begin{eqnarray}
\partial^2_{\chi} (\chi \Psi) +\left [-\frac{1}{4}
+\frac{\rho}{\chi}-\frac{l(l+1)}{\chi^2} \right ] (\chi \Psi)=0\,.
  \label{far wave eq}
\end{eqnarray}
This is a standard Wittaker equation \cite{abramowitz},
 $\partial^2 _\chi W +\left[ -\frac{1}{4}
 + \frac{\rho}{\chi}-\frac{1-\mu^2}{\chi^2} \right ] W=0 $, with
\begin{eqnarray}
& & \hspace{-0.5cm} W=\chi \Psi \,,  \qquad \mu=l+1/2 \,.
 \label{Wittaker parameters}
\end{eqnarray}
The most general solution of this equation is $W=
\chi^{\mu+1/2}e^{-\chi/2}[\alpha \,M(\tilde{a},\tilde{b},\chi)+\beta
U(\tilde{a},\tilde{b},\chi)]$, where $M$ and $U$ are the Wittaker's
functions with $\tilde{a}=1/2+\mu-\rho$ and $\tilde{b}=1+2\mu$. In
terms of the parameters that appear in (\ref{far wave eq}) one has
\begin{eqnarray}
& & \tilde{a}= l+1-\rho \,,
\nonumber \\
 & & \tilde{b}=2l+2 \,.
 \label{Wittaker parameters 2}
\end{eqnarray}
The far-region solution of (\ref{far wave eq}) is then given by
\begin{eqnarray}
 \Psi =\alpha \, \chi^{l} e^{-\chi/2}\,M(\tilde{a},\tilde{b},\chi)
 +\beta \, \chi^{l} e^{-\chi/2}\,U(\tilde{a},\tilde{b},\chi)  \,.
 \label{far field}
\end{eqnarray}

To impose the boundary condition at infinity, we will need to know the
behavior of the far-region solution at the large $\chi=2 \eta r
\rightarrow \infty$ regime. When $\chi \rightarrow +\infty$, one
has $U(\tilde{a},\tilde{b},\chi)\sim \chi^{-\tilde{a}}$ and
$M(\tilde{a},\tilde{b},\chi)\sim \chi^{\tilde{a}-\tilde{b}}
e^{\chi} \Gamma(\tilde{b})/\Gamma(\tilde{a})$ \cite{abramowitz},
and thus for large $\chi$, the far-region solution behaves as
\begin{eqnarray}
\Psi \sim  \alpha \, \frac{\Gamma(2l+2)}{\Gamma(l+1-\rho)}
 \chi^{-1-\rho} e^{\chi/2} +\beta  \chi^{-1+\rho} e^{-\chi/2}
  \,.
 \label{far field large r}
\end{eqnarray}
It is clear that the first term proportional to $e^{+\chi/2}$
represents an ingoing wave, while the second term proportional to
$e^{-\chi/2}$ represents an outgoing wave. At this point we can
finally discuss the second boundary condition of our problem. In
our present experiment, one perturbs the boosted Kerr black string
outside its horizon, and this generates a wave that propagates
both into the horizon and out to infinity. Therefore, at
$r\rightarrow +\infty$, our physical system has only an outgoing
wave, and thus one has to set $\alpha=0$ in (\ref{far field}) and
(\ref{far field large r}). The behavior of the solution at
infinity then boils down to (\ref{bound cond}). We can understand
this boundary condition in an alternative way. As we are already
anticipating with the definition chosen for $\eta^2$ in (\ref{far
wave parameters}) the unstable modes will be those with
$\omega<\kappa$, i.e., whose frequency is smaller than the
potential barrier at infinity. At infinity the wave mode will then
have a {\it real} exponential behavior, $\Psi \sim e^{\pm \chi}$,
and the requirement of regularity at infinity amounts to discard
the solution $e^{+\chi}$ that diverges at infinity. This is done
setting $\alpha=0$.

\subsection{\label{sec:matching} Matching condition. Frequency spectrum}
In this subsection we shall find the real frequencies that are
allowed to propagate in the boosted Kerr black string background,
and we will use the matching of the far and near regions in their
overlapping sector to select the imaginary part of the mode
frequencies.

We begin by asking what are the (real) frequencies that can
propagate in the background of the boosted Kerr black string. To
do so, we shall use a trick \cite{detweiler} that was already
applied with success to solve similar problems in the background
of a Kerr black hole \cite{detweiler}, of a Kerr black string
\cite{cardosobraneinst}, and of a boosted Schwarzschild black
string \cite{CardDiasYosh}. That this method yields indeed the
correct answer has already been confirmed  by a full numerical
checking \footnote{We can do the analysis without making use of
this trick. Indeed, we can find an algebraic equation for the
spectrum of real frequencies and the instability timescale just by
using the second relation of (\ref{far wave parameters}), with
$\rho={\rm Re}[\rho]+i\delta \rho$, and $\delta \rho$ given
by(\ref{delta rho}), and without using (\ref{frequency spectrum}).
However, this trick clarifies the physical nature of the real
frequencies, and yields the right answer, as a full numerical
analysis of the unboosted case confirms
\cite{cardosobraneinst2}.}. We assume for a moment that we have no
black hole, and we ask what are the frequencies that can propagate
in this horizon-free background. In this setup, we are actually
looking for the pure normal modes that can propagate in a geometry
that is horizon-free and that has a reflecting spherical wall
produced by the effective mass of the extra dimension. In this
geometry, the wave solution of the Klein-Gordon equation is
everywhere described by (\ref{far field}), subjected to the
appropriate boundary conditions. The outer boundary condition
requires the presence of only outgoing waves and this implies
$\alpha=0$. The inner boundary condition is that $\Psi$ must be
regular at the origin, $\chi\rightarrow 0$. For small values of
$\chi$, the solution is described by
 $\Psi \sim \beta \, [\Gamma(2l+1)/\Gamma(l+1-\rho)]
(2\eta r)^{-l-1}$ \cite{abramowitz}. So, when $\chi\rightarrow 0$,
the wavefunction $\Psi$ diverges, $r^{-l-1}\rightarrow \infty$. To
have a regular solution there we must then demand that
$\Gamma(l+1-\rho)\rightarrow \infty$. This occurs when the
argument of the gamma function is a non-positive integer,
$\Gamma(-N)=\infty$ with $N=0,1,2,\cdots$. Therefore, the
requirement of regularity imposes the condition $l+1-\rho=-N$.
Since $\rho$ is related to $\omega$, the above regularity
demanding amounts to a natural selection of the allowed
frequencies that can propagate in the geometry.

Now, let us come back to the boosted Kerr black string background,
which differs clearly from the above background due to the
existence of a horizon. In the spirit of
\cite{detweiler,CardDiasAdS}, we expect that the presence of an
horizon induces a small complex imaginary part in the allowed
frequencies, $\omega_i={\rm Im}[\omega]$, that describes the slow
decay of the amplitude of the wave if $\omega_i<0$, or the slowly
growing instability of the mode if $\omega_i>0$. Now, from
(\ref{far wave parameters}), one sees that a frequency $\omega$
with a small imaginary part corresponds to a complex $\rho$ with a
small imaginary part that we will denote by $\delta \rho \equiv
{\rm Im}[\rho] $. Therefore, we anticipate  that in the boosted
Kerr black string one has
\begin{eqnarray}
\rho=(l+1+N)+ i \delta\rho\,,
 \label{frequency spectrum}
\end{eqnarray}
with $N$ being a non-negative integer, and $\delta \rho$ being a
small quantity. In particular, this means that onwards, the
arguments of the Wittaker's function $U(\tilde{a},\tilde{b},\chi)$
previously defined in (\ref{Wittaker parameters 2}) are to be
replaced by
\begin{eqnarray}
& & \hspace{-0.5cm} \tilde{a}=-N- i \delta\rho\,,  \qquad
\tilde{b}=2l+2 \,.
 \label{Wittaker parameters end}
\end{eqnarray}

What we have done so far was to find the spectrum of real
frequencies. To find the imaginary part of the modes, we have to
match the far-region with the near-region. So, we need to find the
small $\chi$ behavior of the far-region solution (\ref{far
field}). The Wittaker's function $U(\tilde{a},\tilde{b},\chi)$ can
be expressed in terms of the Wittaker's function
$M(\tilde{a},\tilde{b},\chi)$ \cite{abramowitz}. Inserting this
relation on
 (\ref{far field}), the far-region solution
can be written as
\begin{eqnarray}
\Psi = \beta \, \chi^{l} e^{-\chi/2}\,\frac{\pi}{\sin(\pi
\tilde{b})} {\biggl [}
\frac{M(\tilde{a},\tilde{b},\chi)}{\Gamma(1+\tilde{a}-\tilde{b})
\Gamma(\tilde{b})}-\chi^{1-\tilde{b}}\,
\frac{M(1+\tilde{a}-\tilde{b},2-\tilde{b},\chi)}{\Gamma(\tilde{a})
\Gamma(2-\tilde{b})} {\biggr ]},
 \label{far field-small x aux}
\end{eqnarray}
with $\tilde{a}$ and $\tilde{b}$ defined in (\ref{Wittaker
parameters end}). Now, we want to find the small $\chi$ behavior
of (\ref{far field-small x aux}), and to extract $\delta \rho$
from the gamma function. This is done in
 (\ref{far small x A1})-(\ref{far small x A3}), yielding for
small $\delta \rho$ and for small $\chi$ the result
\begin{eqnarray}
 \Psi \sim \beta (-1)^N \frac{(2l+1+N)!}{(2l+1)!}
\, (2\eta r)^l + \beta (-1)^{N+1}(2l)! \, N! \, ( i \delta \rho)
\, (2 \eta r)^{-l-1}.
 \label{far field-small x}
\end{eqnarray}

\begin{table}
\begin{center}
\begin{tabular}{|c|c|c|c|c|c|c|c|} \hline
$a$ & \multicolumn{1}{c|}{Re[$\omega$]} & \multicolumn{1}{c|}{
Im[$\omega$]} &\multicolumn{1}{c|}{$4\pi T_H \varpi$}
&\multicolumn{1}{c|}{$a^2(\kappa^2-\omega^2)$}
&\multicolumn{1}{c|}{$\omega a$} &\multicolumn{1}{c|}{$\kappa^2
r_+^2$} \\
\hline
0.198  & $ 0.499885 $   &   $   -3.90 \times 10^{ -14} $  &  $+0.0004$ &   $   4 \times 10^{ -6} $  &  $0.10$ &  $0.23$\\
\hline
 0.199  &$  0.499885 $& $  4.85 \times 10^{ -14} $& $-0.0004 $ &   $   4 \times 10^{ -6} $  &  $0.10$ &  $0.23$\\
\hline
 0.200  & $ 0.499885 $&$     1.36 \times 10^{ -13}  $&$-0. 0012$ &   $  4  \times 10^{ -6} $  &  $0.10$ &  $0.23$\\
\hline
0.300   & $ 0.499885 $&$    7.50 \times 10^{ -12}  $&$- 0. 0894$ &   $  1 \times 10^{ -5} $  &  $0.15$ &  $0.20$\\
\hline
0.400 & $ 0.499885 $& $   1.29  \times 10^{ -11}  $& $- 0.2072$ &   $   2\times 10^{ -5} $  &  $0.20$ &  $0.16$\\
\hline
0.450 & $  0.499885  $&$    1.59  \times 10^{ -11}  $&$- 0.2969$ &   $  2  \times 10^{ -5} $  &  $0.22$ &  $0.13$\\
\hline
0.490 & $  0.499885   $& $   2.00 \times 10^{ -11}  $&$- 0.4316$ &   $  3 \times 10^{ -5} $  &  $0.24$ &  $0.09$\\
\hline
0.499 & $  0.499885   $&  $   2.26 \times 10^{ -11}   $&$- 0.5174$ &   $  3 \times 10^{ -5} $  &  $0.25$ &  $0.07$\\
\hline
\end{tabular}
\end{center}
\caption{\label{tab:var a} Some numerical values of the
instability as we vary the rotation parameter $a$ of a geometry
with $2M=1,\, \sigma={\rm arctanh} {\bigl (}1/\sqrt{2} {\bigr )}$,
and for modes with $\kappa=0.5$ and $l=m=1,\, N=0$.  In the second
and third column we have, respectively, the real part,
Re$[\omega]$, and  the imaginary part, Im$[\omega]$, of the mode
frequency. In the fourth column we present the superradiant factor
$\varpi$ multiplied by $4\pi T_H$. The unstable  modes have
Im$[\omega]>0$ and are those that satisfy simultaneously
$\varpi<0$ and $\omega < \kappa$.  The three last columns show the
values of the quantities that must be small in the regime of
validity of our results, namely,  $a^2(\kappa^2-\omega^2) \ll 1$,
$\omega a \ll 1 $  and  $\kappa^2 r_+^2 \ll 1 $. Note that the
condition $\omega^2 r_+^2 \ll 1 $ is automatically satisfied when
$\kappa^2 r_+^2 \ll 1 $, since the unstable modes we are dealing
with are those that have $\omega< \kappa$.}
\end{table}

\begin{table}
\begin{center}
\begin{tabular}{|c|c|c|c|c|c|c|c|} \hline
$\kappa$ & \multicolumn{1}{c|}{Re[$\omega$]} &
\multicolumn{1}{c|}{ Im[$\omega$]} &\multicolumn{1}{c|}{$4\pi T_H
\varpi$} &\multicolumn{1}{c|}{$a^2(\kappa^2-\omega^2)$}
&\multicolumn{1}{c|}{$\omega a$} &\multicolumn{1}{c|}{$\kappa^2
r_+^2$}
 \\
\hline
0.01  &  $ 0.00999  $   &   $   1.67 \times 10^{ -26} $  &  $-0.285$  &   $  2 \times 10^{ -12} $  &  $0.004$ &  $ 0.0007$\\
\hline
 0.10  & $0.09999 $   &   $   1.38 \times 10^{ -17} $  &  $-0.259$ &   $  2 \times 10^{ -8} $  &  $0.04$ &  $0.01$\\
\hline
 0.30  &$ 0.29998$& $  1.74 \times 10^{ -13} $& $-0.201 $ &   $  2 \times 10^{ -6} $  &  $0.10$ &  $0.07$\\
\hline
 0.50  & $ 0.49988 $&$      1.04 \times 10^{ -11}  $&$-0.142 $ &   $   1.4 \times 10^{ -5} $  &  $0.18$ &  $0.18$\\
\hline
0.70   & $ 0.69968 $&$    1.11 \times 10^{ -10}  $&$- 0.084 $ &   $  5 \times 10^{ -5} $  &  $0.25$ &  $0.36$\\
\hline
0.90 & $ 0.89932  $& $   3.10  \times 10^{ -10}  $& $- 0.026$ &   $  2 \times 10^{ -4} $  &  $0.31$ &  $0.59$\\
\hline
1.00  & $   0.99907$&$   -  9.90 \times 10^{ -11}  $&$+ 0.003$ &   $   2\times 10^{ -4} $  &  $0.35$ &  $0.73$\\
\hline
\end{tabular}
\end{center}
\caption{\label{tab:var k} Some numerical values of the
instability for a geometry with $2M=1,\, a=0.35,\, \sigma={\rm
arctanh} {\bigl (}1/\sqrt{2} {\bigr )}$, and for modes with
$l=m=1,\, N=0$ and several values of the KK momentum $\kappa$. In
the second and third column we have, respectively, the real part,
Re$[\omega]$, and  the imaginary part, Im$[\omega]$, of the mode
frequency. In the fourth column we present the superradiant factor
$\varpi$ multiplied by $4\pi T_H$. The unstable  modes have
Im$[\omega]>0$ and are those that satisfy simultaneously
$\varpi<0$ and $\omega < \kappa$.  The three last columns show the
values of the quantities that must be small in the regime of
validity of our results. }
\end{table}

The quantity $\delta \rho$ cannot take any value. Its allowed
values are selected by requiring a match between the near-region
solution (\ref{near field large r}) and the far-region
solution (\ref{far field-small x}). So, the allowed values of
$\delta \rho$ are those that satisfy the matching condition
 \begin{eqnarray}
 - i \delta \rho \frac{(2l)!(2l+1)! N!}
 {(2l+N+1)!(2\eta )^{2l+1}}
 =(r_+-r_-)^{2l+1}
\frac{\Gamma(l+1)}{\Gamma(2l+1)}\frac{\Gamma(-2l-1)}{\Gamma(-l)}
\frac{\Gamma(l+1- i \,2\varpi)}{\Gamma(-l- i \,2\varpi)} \,.
 \label{Matching Instab}
\end{eqnarray}
Use of the gamma function relations
(\ref{gamma values 3}) yields
\begin{eqnarray}
\delta \rho  = -2  \varpi[2\eta
(r_+-r_-)]^{2l+1}\frac{(2l+1+N)!}{N!}\,\left [
\frac{l!}{(2l)!(2l+1)!}\right ]^2\prod_{\jmath=1}^l
(\jmath^2+4\varpi ^2)\,,
 \label{delta rho}
\end{eqnarray}
for $l\geq 1$, while
for $l=0$ one gets
 \be \delta \rho =-4\eta (r_+-r_-) (N+1)
\varpi\,.\ee
 Condition (\ref{frequency spectrum}) together with
  (\ref{far wave parameters}) leads to
 \beq
\frac{(r_+ + r_-)(\omega \sinh \sigma-\kappa \cosh
\sigma)^2}{2\sqrt{\kappa^2-\omega ^2}}=l+N+1+i \delta \rho\,,
\label{condfinal} \eeq
 with $\delta \rho$ given by (\ref{delta rho}).
This is an algebraic equation for the characteristic values of the
frequency $\omega$. All these values must be consistent with the
assumptions made, namely,  (\ref{eigenvalues}), (\ref{near wave
eq}) and (\ref{far wave eq aux}) are valid only for
$a^2(\kappa^2-\omega^2) \ll 1$, $\omega a \ll 1 $  and $\kappa^2
r_+^2 \ll 1 $.  If $\omega$ has a positive imaginary part, then
the mode is unstable. Indeed, the field has the time dependence
$e^{-i\omega t}$, so a positive imaginary part for $\omega$ means
the amplitude grows exponentially as time goes by. We have indeed
found unstable modes. Representative elements of this class of
modes are displayed in Table \ref{tab:var a} (where we fix the
Kaluza-Klein momentum $\kappa$ of the mode and vary the rotation
$a$ of the geometry), and in Table \ref{tab:var k} (where we fix
$a$ and vary $\kappa$). As a consistency check, in Appendix
\ref{sec:A2}, we show that when we turn off the boost,
(\ref{condfinal})  yields the result found originally by Detweiler
\cite{detweiler}.

We discuss the results in section \ref{discussion}.

\section{\label{sec:Hi dim}Stability analysis of the
boosted Myers-Perry black string}

The  boosted Myers-Perry black string can also be constructed by
adding a flat direction $z$ to the Myers-Perry black hole
\cite{myersperry} \footnote{The Myers-Perry black hole lives in a
background with $D\geq 5$ spacetime dimensions. The corresponding
black string lives in a $(D+1)$-dimensional background.}, and then
applying a Lorentz boost to it, $dt\rightarrow \cosh\sigma
dt+\sinh\sigma dz$ and $dz\rightarrow \sinh\sigma dt+\cosh\sigma
dz$. The geometry of a boosted Myers-Perry black string geometry
is then (for our purposes it is enough to consider the case with
rotation in a single plane parameterized by $\phi$)
\begin{eqnarray}
 ds^2&=&-\left( 1-\frac{2M r^{1-n} \cosh^2\sigma }{\Sigma}\right) dt^2
 + \frac{2M r^{1-n} \sinh(2\sigma) }{\Sigma}
 \,dt dz + \left( 1+\frac{2M r^{1-n} \sinh^2\sigma }{\Sigma}\right)dz^2
 +\frac{\Sigma}{\Delta}dr^2
 +\Sigma d\theta^2  \nonumber \\
 & & \hspace{-0.5cm} +\frac{(r^2+a^2)^2-\Delta a^2
\sin^2\theta}{\Sigma}\,\sin^2\theta d\phi^2
  -\frac{4M r^{1-n} \cosh\sigma }{\Sigma}\,a\sin^2\theta dt d\phi
  -\frac{4M r^{1-n} \sinh\sigma }{\Sigma-2Mr^{1-n}}\,a\sin^2\theta dz d\phi
  +r^2\cos^2\theta d\Omega_n^2  \,, \nonumber \\
 & &
   \label{metric Hi dim}
\end{eqnarray}
where $n=D-4$, $d\Omega_n^2$ describes the line element of a unit
$n$-sphere, and
\begin{eqnarray}
\Delta = r^2+ a^2 -2Mr^{1-n} \,, \qquad  \Sigma = r^2+
a^2\cos^2\theta \,.
   \label{metric parameters Hi dim}
\end{eqnarray}

Under the separation ansatz
 \begin{eqnarray}
  \Phi= e^{-i \omega t}e^{i m \phi} e^{-i \kappa z}
  S^m _l(\theta)\Psi(r) Y(\Omega) \,, \label{ansatz hiDIM}
 \end{eqnarray}
the Klein-Gordon equation yields the following angular and radial
wave equations for $S^m _l(\theta)$ and $\Psi(r)$,
\begin{eqnarray}
 & &\hspace{-0.5cm} \frac{1}{\sin \theta \cos^n \theta}
\partial_{\theta}\left ( \sin \theta \cos^n \theta \partial_{\theta}
S^m _l \right )  + \left [  a^2 (\omega^2-\kappa^2) \cos^2
\theta-\frac{m^2}{\sin ^2{\theta}}+\lambda_{lm} -\frac{j(j+n-1)}{\cos^2 \theta}
\right ]S^m _l =0\,, \label{separation ang hiDIM}
\\
& &\hspace{-0.5cm} \frac{\Delta}{r^n}  \partial_r
\left ( r^n\Delta \partial_r \Psi \right ) +U \Psi
=0\,,
 \label{separation rad hiDIM}
\end{eqnarray}
where $\lambda_{lm}$ is the ($n$-dimensional) separation constant given by
$\lambda_{lm}=l(l+1+n)+{\cal O}{\bigl ( }a^2(\omega^2-\kappa^2){\bigr )}$, and
\begin{eqnarray}
U&=& - \Delta {\biggl [}  \kappa^2 r^2 +a^2\omega^2-2\omega m a
\cosh\sigma +\lambda_{lm} +j(j+n-1)\,\frac{a^2}{r^2} {\biggr ]} +
{\biggl [}  {\bigl [}\omega(r^2+a^2)-m a \cosh\sigma {\bigr ]}^2
\nonumber \\
& &
 + 2M r^{1-n}(r^2+a^2)\cosh^2\sigma
 {\bigl [} \omega-\kappa\tanh\sigma {\bigr ]}^2  -2Mr^{1-n}(r^2+a^2)\omega^2
 -m^2a^2 \sinh^2\sigma +4\kappa m a M r^{1-n} \sinh\sigma {\biggr ]}\,,
 \label{potential U hiDIM}
\end{eqnarray}
In (\ref{separation ang hiDIM}) and (\ref{potential U hiDIM}), the
integer $m = 0, \pm 1, \pm 2,\cdots$ comes from separation of the
angle describing the azimuthal dependence of the perturbations
around the symmetry axis [see (\ref{ansatz hiDIM})]. The terms
dependent on $n$ and the parameter $j = 0, 1, 2,\cdots$ are the
eigenvalues of the hyperspherical harmonics on the $n$-sphere,
which are given by $-j(j + n - 1)$ \cite{Berti:2005gp}. The
integer $l$ is constrained to satisfy the condition $l \geq (j
+|m|)$. Note that (\ref{separation ang hiDIM})-(\ref{potential U
hiDIM}) reduce to (\ref{separation ang})-(\ref{separation rad})
when we set simultaneously $n=0$ and $j=0$.

The boundary conditions are only ingoing flux at the
horizon and only outgoing waves at infinity,
 \begin{equation}
\Psi(r)\sim
 \left\{
\begin{array}{ll}
(r-r_+)^{-i \varpi }=e^{-i \varpi  \ln (r-r_+)}\,, \qquad  {\rm
as} \quad r\rightarrow r_+ \,, \\
r^{-(n+2)/2}\,e^{+i
\sqrt{\omega^2-\kappa^2} \,r}\,, \qquad   {\rm as} \quad
r\rightarrow \infty \,,
\end{array}
\right.
\label{bound cond hiDIM}
\end{equation}
where
\begin{eqnarray}
\varpi = \frac{r_+(r_+^2+a^2)\cosh\sigma}{(n+2)r_+^2+(n-1)a^2}\,
\left(\omega-\omega_{sup} \right) \,,
 \label{lambda hiDIM}
\end{eqnarray}
with $\omega_{\rm sup}$ defined in (\ref{wsup}).

\begin{figure}[b]
\begin{center}
{\includegraphics{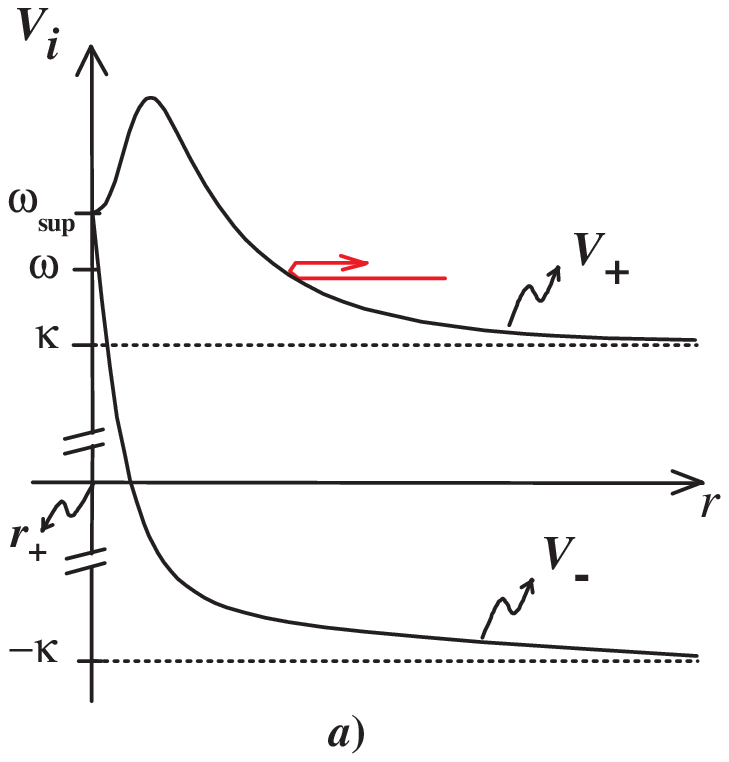}\qquad
\includegraphics{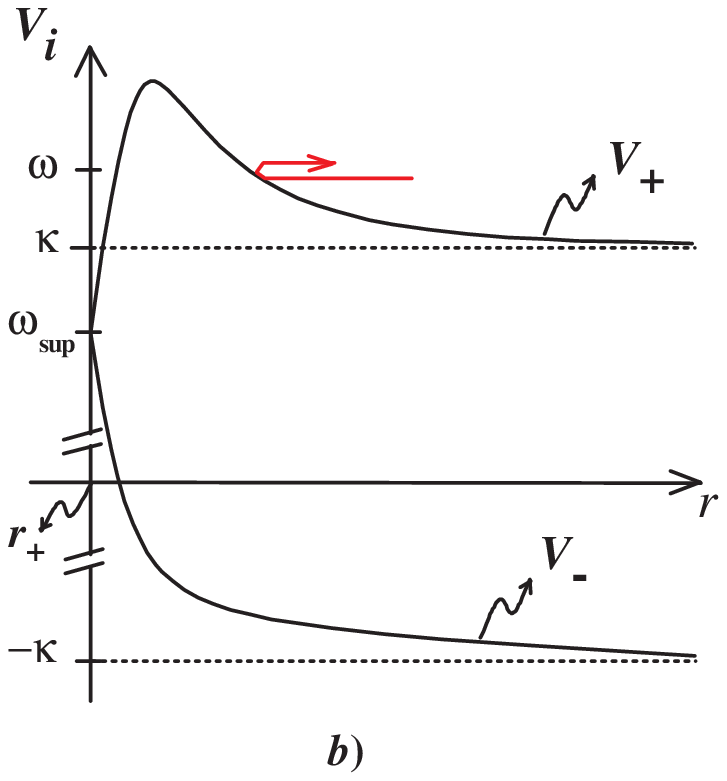}}
\end{center}
\caption{ {\bf a)} General qualitative shape of the potentials
$V_+$ and $V_-$ for a boosted Myers-Perry black string. In this
case the geometry is always stable, but superradiant scattering is
possible. An example of data that yields this kind of potentials
is $(2M=1\,,a=0.9\,,
\tanh\sigma=1/\sqrt{2}\,,\kappa=1.5\,,l=m=1,\, j=0, \, n=1)$.
There are no bound state modes since $V_+$ approaches $\kappa$ at
infinity from above. This absence of bound states for $n\geq 1$,
as oppose to the $n=0$ case, annihilates the possibility of
existing an instability for $n\geq 1$. Since $\omega< \omega_{\rm
sup}$, superradiant scattering is possible. \\
{\bf b)} Qualitative shape of the potentials $V_+$ and $V_-$ for a
Myers-Perry black string. In this case the geometry is stable, and
moreover superradiant scattering is not possible. An example of
data that yields this kind of potentials is $(2M=1\,,a=0.5\,,
\tanh\sigma=1/\sqrt{2}\,,\kappa=2\,,l=m=1,\, j=0, \, n=1)$. There
are no bound state modes and superradiant scattering is not
possible (since $\omega> \omega_{\rm sup}$).  }
 \label{fig:potential HiD}
\end{figure}

Defining the tortoise coordinate $r_{\ast}$ and a new wavefunction
$\chi$ as,
\begin{eqnarray}
d r_{\ast}= \frac{r_+^2+a^2}{\Delta} \, dr\,, \qquad \Psi={\bigl [}
(r_+^2+a^2) r^n{\bigr ]}^{-1/2}\chi \,,
 \label{Schr aux hiDIM}
\end{eqnarray}
the radial wave equation (\ref{separation rad hiDIM}) can be
written in the form of an effective Schr$\ddot{\rm o}$dinger
equation,
 \begin{eqnarray}
  \partial_{r_{\ast}}^2\chi-V\chi=0 \,,
  \label{Schr hiDIM}
 \end{eqnarray}
 with
  \begin{eqnarray}
  V=-\frac{U}{(r^2+a^2)^2}+G^2+\partial_{r_{\ast}}G \,, \qquad
  G= \frac{1}{2} \partial_{r_{\ast}} \ln [(r_+^2+a^2) r^n] \,.
  \label{Schr potential hiDIM}
 \end{eqnarray}
For  $a^2(\omega^2-\kappa^2) \ll 1$ one has $\lambda_{lm}\approx l(l+1+n)$
and V is then a quadratic function of $\omega$ that can be factorized as
 \begin{eqnarray}
  V=-\gamma (\omega-V_+) (\omega-V_-)  \,, \label{V hiDIM}
 \end{eqnarray}
with
 \begin{eqnarray}
& & \gamma= \frac{ r^{1-n}}{(r^2+a^2)^2}\,{\bigl [} r^{1+n}(r^2+a^2)
+2M (r^2\sinh^2\sigma +a^2\cosh^2\sigma) {\bigr ]} >0 \,, \nonumber \\
& &
 V_{\pm}= - \frac{V_1}{2\gamma} \pm \sqrt{ \left (
 \frac{V_1}{2\gamma} \right )^2 -\frac{V_0}{\gamma} }\,,
  \label{V+V- hiDIM}
 \end{eqnarray}
and
 \begin{eqnarray}
 V_0 &=& -\frac{j(j+n-1)a^2 \Delta }{r^2 (r^2+a^2)^2}
 +\frac{-[\kappa^2 r^2+ l(l+1+n)]\Delta +m^2 a^2 +4\kappa a m M r^{1-n}
 \sinh\sigma }{(r^2+a^2)^2}
 + \frac{2M \kappa^2  r^{1-n} \sinh^2\sigma }{r^2+a^2} \nonumber \\
& & -\frac{r^{-n-2} \Delta }{4 (r^2+a^2)^4} \, {\biggl [} n(n-2)a^6 r^n+ (n+2) r^5
{\bigl [} 2M(n+2)+n r^{1+n}   {\bigr ]}  + a^4 r{\bigl [} 2M n^2+ (3n^2-2n+4)r^{1+n}
 {\bigr ]} \nonumber \\
& &
  \hspace{2.4cm}+  a^2 r^3 {\bigl [} 4M (n^2+2n-4)+ (3n^2+2n+4)r^{1+n}  {\bigr ]}
  {\biggr ]}   \,,
    \nonumber \\
 V_1 &=& - \frac{4Mr^{1-n}\cosh\sigma {\bigl [} m a+
 \kappa \sinh\sigma (r^2+a^2){\bigr ]} }{(r^2+a^2)^2}\,, \label{V0V1 hiDIM}
 \end{eqnarray}
Important features of $\gamma$ and $V_{\pm}$ are their asymptotic
values at the horizon and infinity which are given by
 \begin{eqnarray}
 & &  \lim_{r\rightarrow r_+} \gamma= \cosh^2 \sigma \,, \qquad
  \lim_{r\rightarrow \infty} \gamma= 1\,, \nonumber \\
& &
  \lim_{r\rightarrow r_+} V_{\pm}= \omega_{\rm sup}\ \,, \qquad
  \lim_{r\rightarrow \infty} V_{\pm}= \pm |\kappa| \,. \label{lim V hiDIM}
 \end{eqnarray}

Note that when $n=0$ (which implies $j=0$), the relations of this
section yield the potentials discussed in section \ref{sec:schr
4D}. In this case, we have already analyzed the potentials
$V_{\pm}$ in Figs. \ref{fig:potential 4D}.(a) and
\ref{fig:potential 4D}.(b). For $n\geq 1$, the relevant features
of the potentials are independent of $n$, and significantly
different from the $n=0$ case. The only two kinds of potentials
that can occur for $n\geq 1$  are drawn in Figs.
\ref{fig:potential HiD}.(a) and \ref{fig:potential HiD}.(b).
Superradiant modes are allowed (Fig. \ref{fig:potential HiD}.(a)),
but the KK momentum is never able to generate a potential well,
i.e., bound states, and thus the superradiant instability does not
develop. When we set the boost to zero, $\sigma=0$, our results
and graphs reduce to the ones found in
\cite{cardosobraneinst2,typo}. As pointed out in
\cite{cardosobraneinst2}, this absence of instability in the
$n\geq 0$ case seems to be related with the fact that, at least
for $n=1$, there are no stable circular orbits \cite{frolov}.

We end this section with a remark on the value of the boost angle
for $n\geq 1$. As is being assumed in this section, if black rings
exist in six and higher dimensions, their infinite radius limit
will  be a boosted Myers-Perry black string. However,  the
balanced rings need not give boosted strings with
$\tanh\sigma=1/\sqrt{2}$ (the value for the Kerr black string).
Presumably, but we do not know for sure, the balanced rings give
pressureless boosted black strings. To find the relevant value for
$\sigma$ in the higher dimensional case one would have to solve
the equation $T_{zz}=0$, where $T_{zz}$ will be defined in
(\ref{myers tensor}). The result for $\sigma$ may therefore depend
on the dimension of the spacetime. We have tested several values
of $\sigma$ and verified that we always get potentials like the
ones plotted in Fig. \ref{fig:potential HiD}, that show no
evidence of superradiant instability.

\section{\label{discussion}Discussion of the results.
The instability endpoint. Bound mechanism for the rotation}

We have shown that the 5D large radius doubly spinning black rings
are unstable against  superradiant bound modes. The most
physically intuitive method to identify the nature of the
instability is to write the radial wave equation as a
Schr$\ddot{\rm o}$dinger equation and to look to the factorized
Schr$\ddot{\rm o}$dinger potentials, as was done in section
\ref{sec:schr 4D}. When the instability is present the potentials
are like the ones drawn in Fig. \ref{fig:potential 4D}.(a). The
two ingredients necessary for the activation of the instability
are the existence of a potential well in $V_+$ that can trap bound
states, and that superradiance is present. In the boosted Kerr
black string ($n=0$) these two ingredients can be simultaneously
present. Therefore, the large radius 5D black rings are unstable
against this mechanism. We also note that there are in addition
damped modes which are bound states that do not suffer
superradiance and die off through the tunneling to the horizon, as
shown in Fig. \ref{fig:potential 4D}.(b).

To find the frequency spectrum of the unstable modes and the
timescale of the process we have used the matched asymptotic
method.  The main features of the instability found for the large
radius black ring solution (the boosted Kerr black string) can be
summarized as follows:

\noindent (i) In Table \ref{tab:var a} we show the evolution of
the instability when the rotation $a$ of the boosted Kerr black
string increases, while keeping all other parameters of the
geometry and angular parameters of the mode fixed. We see that the
instability is robust and it gets stronger (the timescale
$\tau\sim 1/{\rm Im}[\omega]$ decreases) when the rotation $a$
increases. There is natural explanation for this behavior. The
instability is triggered by superradiance which is present only
when the background rotates. Therefore we expect the superradiance
and thus the instability to be stronger when the rotation is
bigger: in Table \ref{tab:var a}, one sees that as
 $a$ increases so does $|\varpi|$ and ${\rm Im}[\omega]$.

\noindent (ii) Climbing Table  \ref{tab:var k} from the bottom to
the top we verify that a transition between an unstable situation
to a stable one occurs between the second and first line. This
happens because below a critical rotation the supperradiant
condition ${\rm Re}[[\omega]< \kappa \tanh \sigma+m \Omega_{\phi}
\cosh^{-2}\sigma$ is no longer valid, and it clearly identifies
superradiance as the mechanism responsible for the instability.

\noindent (iii) In Table \ref{tab:var k} we show the evolution of
the instability when the KK momentum $\kappa$ of the mode
increases, while keeping all other parameters of the geometry and
angular parameters of the mode fixed. In general, the instability
is robust and it gets stronger when $\kappa$ increases. This
behavior can be understood by noting that the potential well of
$V_+$ in Fig. \ref{fig:potential 4D}.(a)  gets more deep when
$\kappa$ increases. Thus, the bound modes get more efficiently
trapped in the well.

\noindent (iv) However, when $\kappa$ increases above a critical
value the supperradiant condition ${\rm Re}[\omega]< \kappa \tanh
\sigma+m \Omega_{\phi} \cosh^{-2}\sigma$ is no longer valid and
the instability stops. This situation can be seen in the
transition between the two last lines of Table  \ref{tab:var k}.
Again, this is a strong evidence that  superradiance is the
 mechanism responsible for the instability.

\noindent (v) The most unstable mode we have found has a timescale
$\tau\sim 1/{\rm Im}[\omega]\sim 10^{7}$, frequency ${\rm
Re}[\omega]=1.84398$, KK momentum $\kappa=1.85$ and $l=m=1$. This
mode is present in the geometry $2M=1$, $a=0.499999$ and
$\tanh\sigma=1/\sqrt{2}$. The most unstable mode when there is no
boost, $\sigma=0$, has a timescale $\tau \sim 10^{8}$
\cite{cardosobraneinst2}. Therefore the boost can increase
considerably the strength of the instability.

\noindent (vi)  When we set the rotation  parameter to zero, $a=0$
(boosted Schwarzschild black string), we always get potentials of
the form represented in Fig. \ref{fig:potential 4D}.(b), for any
$M\,,\sigma$ and $\kappa\,,l\,,m$. In particular this means that
the bound states of the boosted black string are always damped
modes since the  superradiant condition (\ref{super cond}) is
never satisfied. This was first noticed in
\cite{NozMaeda,CardDiasYosh}, where it was shown that superradiant
scattering is never possible in a boosted  Schwarzschild black
string, although the particle analogue of this phenomena -- the
Penrose process -- can occur.

We have shown that infinite radius black rings have unstable
scalars modes. However, we also expect gravitational unstable
modes of the same kind to be present. Indeed, metric modes will
also have KK momentum that provides the potential well that allows
the existence of bound states, and it is well known that
superradiant scattering also occurs for gravitational
perturbations \cite{zelmisnerunruhbekenstein,staro1}. We thus have
the ingredients for the development of the instability. Moreover
this gravitational instability will be considerably stronger than
the scalar one. Indeed, the maximum superradiant amplification
after each scattering is only $2 \%$ in the scalar case but  it
reaches $138\%$  in the gravitational case
\cite{PressTeuAmplifFact}.

\vskip 0.2cm

A question that naturally arises whenever one has an instability
is what is its endpoint. Quite often, this problem has not a clear
answer, the most paradigmatic example being the endpoint of the
Gregory-Laflamme instability. The endpoint of the superradiant
instability is however more predictable. Indeed, the amplitude of
an unstable mode gets amplified during each scattering. The total
energy of the system is conserved because the rotational energy of
the boosted Kerr black string decreases during the process. Thus,
the rotation $a$ decreases as the instability evolves until it
reaches the critical minimum value for which the superradiant
factor $\varpi$ vanishes (see the second item above). At this
point the instability stops, and we have a black string with lower
rotation than the initial one and some rotating radiation around
it trapped in the potential well (that eventually escapes to
infinity). This would be the full story if we only had one
unstable mode with a specific value of $\kappa\,,l\,,m$. However,
mode perturbations with several different values $\kappa\,,l\,,m$
are spontaneously generated in the vicinity of the horizon. And
the smaller the value $\kappa$ of the unstable mode is, the
smaller is the minimum critical value of rotation, $a_{\rm crit}$,
for which superradiance (and the instability) ceases to operate,
as indicated in Table \ref{tab:a crit}. Therefore we conclude that
the several unstable modes will effectively spin down completely
the boosted Kerr black string until its rotation along $S^2$
vanishes, and we are left with a boosted black string. Note
however that table \ref{tab:a crit} also tells us that as the
critical rotation decreases, the timescale of the unstable mode
also gets significantly weaker. So the final stage of the complete
spin down will take in practice a long time. Summarizing,  a large
radius doubly spinning black ring with rotation both along $S^1$
and $S^2$ will decay into a large radius black ring that rotates
only along $S^1$.
\begin{table}
\begin{center}
\begin{tabular}{|c|c|c|c|c|c|c|c|} \hline
$\kappa$&\multicolumn{1}{c|}{$a_{\rm crit}$}
&\multicolumn{1}{c|}{$4\pi T_H \varpi$}
& \multicolumn{1}{c|}{Re[$\omega$]}
& \multicolumn{1}{c|}{
Im[$\omega$]} &\multicolumn{1}{c|}{$a^2(\kappa^2-\omega^2)$}
&\multicolumn{1}{c|}{$\omega a$} &\multicolumn{1}{c|}{$\kappa^2 r_+^2$}
 \\
\hline
0.50 &  $ \sim 0.1985  $   &  $\sim 0^+ $ &  $ 0.49988  $   &   $   4.79 \times 10^{ -15} $    &   $  5 \times 10^{ -6} $  &  $0.099$ &  $ 0.23$\\
\hline
 0.10 &  $ \sim 0.0414  $   &  $\sim 0^+ $ & $0.09999 $   &   $   2.44 \times 10^{ -21} $   &   $  3 \times 10^{ -10} $  &  $0.004$ &  $0.01$\\
\hline
 0.01  &  $ \sim 0.0043  $  &  $\sim 0^+ $ &$ 0.00999$& $  6.28 \times 10^{ -30} $ &   $  3 \times 10^{ -16} $  &  $0.00004$ &  $0.0001$\\
\hline
  \end{tabular}
\end{center}
\caption{\label{tab:a crit} The critical value for the rotation
parameter, $a_{\rm crit}$, for which the superradiant factor
vanishes, $\varpi \sim 0$, for three different values of $\kappa$.
 The other values not specified in the table are $2M=1\,,
 \sigma={\rm arctanh} {\bigl (}1/\sqrt{2} {\bigr )}$,
 and  $l=m=1,\, N=0$.  For $a< a_{\rm crit}$, the instability
 is not active because superradiance is no longer present.}
\end{table}

We can ask if we can extend these results for black rings with
finite radius. For example it would be important to find if the potential
barrier of height $\kappa$ at infinity is also present in small radius rings.
 To fully address this issue would require finding the line element of the
doubly spinning black ring. Then, we would have to perform a full
numerical analysis, since even when one of the rotations vanishes,
the Klein-Gordon equation does not separate \cite{CardDiasYosh}.
However, with the present knowledge, we can still address partially
this question. Indeed, we can keep the approximation
in which we use the boosted Kerr black string as a toy-model for
the doubly spinning black ring, but now we compactify the
transverse direction, $z\sim z+2\pi R$, and we decrease slowly
$R$. In this case the KK momentum of a mode that propagates in the
geometry gets quantized and can take only discrete values:
$\kappa= \frac{k}{R}$, with $k$ being an integer. This
discretization of $\kappa$ has important consequences. Indeed, now
there is a minimum value for $\kappa$ and thus there is a
non-vanishing minimum for the critical rotation  $a_{\rm crit}$
below which there is no superradiance and no instability. So we
expect that a finite radius doubly spinning black ring will loose
angular momentum along $S^2$ and will stabilize into a doubly
spinning black ring with small, but non-vanishing, rotation along
$S^2$. The superradiant instability will effectively impose an
upper bound on the rotation along $S^2$. We can estimate the value
of this upper bound. To stabilize the system superradiance cannot
be present, i.e., we must have $\varpi \geq 0$, with $\varpi$
defined in (\ref{superrad factor}). As we see in Tables
\ref{tab:var a}-\ref{tab:a crit}, the unstable modes are waves
with ${\rm Re}[\omega]\sim \kappa$. Therefore, in (\ref{superrad
factor}) we can replace the frequency by the KK momentum. Use of
$\kappa= k / R$ yields that superradiance and hence the
instability will be absent for black rings that satisfy the
relation
\begin{eqnarray}
  R< k \cosh\sigma(1-\tanh\sigma) \frac{r_+^2+a^2}{m a} \,. \label{endpoint}
\end{eqnarray}
We can cast this relation in a more interesting form that gives
an uper bound for the conserved angular momentum of the ring along $S^2$.
To do so, we shall use the model of the black ring recently proposed by
Hovdebo and Myers \cite{boost}, with a small extension to accommodate
a small angular momentum  along $S^2$.  The endpoint of the instability
will have a small rotation along $\phi$. It is then sufficient to work in the small
$a$ regime, for which one has   $r_+^2+a^2 \sim r_+^2$.
The Hovdebo-Myers model assumes that the main features of a finite radius
black ring can be reproduced by taking a loop of string with radius
$R$  and with ADM-like stress tensor given by  \cite{MyersTensor},
\begin{eqnarray}
T_{ab} =\frac{1}{16\pi} \oint d\Omega_2 \hat{r}^2 n^i
{\bigl [} \eta_{ab} \left( \partial_i h^c_{\: c} +\partial_i h^j_{\: j}
- \partial_j h^j_{\: i} \right ) -\partial_i h_{ab} {\bigr ]} \,,
\label{myers tensor}
 \end{eqnarray}
where $d \Omega_2$ is the line element of a 2-sphere,
$n^i$ is a radial unit-vector in the transverse subspace,
$\eta_{\mu\nu}$ is the flat space metric, and
$h_{\mu\nu} = g_{\mu\nu} - \eta_{\mu\nu}$ is the deviation of the
asymptotic metric from flat space. The index labels $a, b, c
\in \{t, z\}$, while $i, j$ run over the
transverse directions. To apply this formula, the asymptotic
metric must approach that of flat space in Cartesian coordinates.
In the  $a \ll r_+$ limit, this is accomplished by applying the
coordinate transformation $r = \hat{r}[1 + r_+/ (2  \hat{r})]$ to
(\ref{metric})  \cite{boost}. Using (\ref{myers tensor}), one finds that
the total energy, angular momentum along $S^1$, and
angular momentum along $S^2$
are respectively given by
 \begin{eqnarray}
\hat{M} &=& 2\pi R \, T_{tt}=\frac{\pi}{2}R r_+ (\cosh^2 \sigma +1)\,, \nonumber \\
\hat{J}_{\psi} &=& 2\pi R^2 \,T_{tz}=\frac{\pi}{2}\, R^2 r_+ \cosh \sigma  \sinh \sigma\,, \nonumber \\
\hat{J}_{\phi} &=& 2\pi R \, T_{t\phi}=\pi  R r_+ a \cosh \sigma \,.
\label{conserv ADM}
 \end{eqnarray}
Using these conserved charges we can rewrite (\ref{endpoint}) as
    \begin{eqnarray}
\hat{J}_{\phi} \lesssim \frac{k}{m} \left( \frac{r_+}{R} \right)^2
 \left[ R \left( \hat{M}-\frac{\pi r_+  R}{2} \right)- \hat{J}_{\psi}  \right] \,.
\label{endpoint 2}
 \end{eqnarray}
 Here, we confirm that for large radius rings the superradiant instability
 imposes $\hat{J}_{\phi} \rightarrow 0$.
 Note that $(\ref{endpoint 2})$ also suggests that
 $\hat{J}_{\psi}< R \left( \hat{M}-\frac{\pi r_+  R}{2} \right)$. That is, doubly spinning
black rings will also have an upper bound for the angular momentum
along $S^1$, contrary to what happens with the black ring rotating
only along $S^1$. This  upper bound on $\hat{J}_{\psi}$ is present
in dipole black rings rotating along $S^1$, and this is an indication
that a similar case should occur in the doubly spinning black rings \cite{BRDipole}.
Relation (\ref{endpoint 2}) does not depend only on the conserved charges.
The string model depends on four independent parameters,
$R$, $r_+$, $\sigma$ and $a$, which correspond respectively to the
size of the loop, thickness of the loop, tangential boost velocity along the loop,
and rotation transverse to the loop \cite{boost}. The system (\ref{conserv ADM})
gives only three relations between the above four parameters
for a fixed configuration of conserved charges. The fourth parameter is fixed
by demanding that the ring acquires a configuration that maximizes its entropy
\cite{boost}.  This last quantity is computed using  the area per
unit length of the horizon, $A_H$, defined in
(\ref{boostedbs TVA}).  One finds that the entropy is given by
$S=2\pi R A_H/4 =2\pi^2 R (r_+^2+a^2)\cosh \sigma
\sim 2\pi^2 R r_+^2 \cosh \sigma$, where the last approximation is valid
in the small rotation regime.
In the model of \cite{boost}, it was found that this entropy is maximized for
$\sigma=\sigma_{\rm max}\sim 0.709$. Use of  (\ref{conserv ADM}) now allows
us to write both $r_+$ and $R$ as a function of the conserved charges.
Then, the upper bound for the ADM angular momentum along $S^2$ can
finally be written as
 \begin{eqnarray}
\hat{J}_{\phi} \lesssim 10^{-3} \pi
\,\left( \frac{\hat{M}^2}{\hat{J}_{\psi} } \right)^3.
\label{endpoint 3}
 \end{eqnarray}
 where we have set $k / m\sim 1$, since we do not expect high
multipoles to be excited in a slowly rotating ring along $\phi$.
This is one of the main results of this paper. Basically, it is
telling us that the instability  provides a dynamical mechanism
that bounds the rotation of the doubly spinning black ring along
$S^2$.

It is important to note that the demand of a regular horizon in
the yet to be found doubly spinning black ring will introduce a
different upper bound for the rotation. The question is then if
the upper bound imposed dynamically by the instability is
or not  smaller than the Kerr-like upper bound imposed by the existence of a
regular horizon. To address this issue heuristically, and since
$\hat{J}_{\psi}$ plays a minor role in this instability, we make
use of the black ring that rotates along $S^2$ (see section
\ref{brS2} \cite{Figueras}).
The comparison is more clear when the relations are written in terms of
the parameters $a$ and $M$.
We first rewrite  (\ref{endpoint 2}) in terms of $a$ yielding $a/M \lesssim  2M /R$,
in the limit $\sigma \sim 0$ and $a\ll r_+$.  Next, we use the first
relation of (\ref{coord transf S2}) in the first inequality of
(\ref{equil cond S2}) to write the horizon bound as  $a/M <  1$
(as expected, since in the large radius limit this is the Kerr
bound). Therefore, for $2M /R \lesssim1$, the instability plays an
important role in the evolution of the solution, i.e., the
instability will afflict large radius black rings but not small
radius black rings.


\vskip 0.2cm

Recently it has been conjectured the existence of higher
dimensional black rings with horizon topology $S^1 \times
S^{n+2}$, with $n=D-4> 0$ \cite{EmpMyersPriv}. Evidence for this
existence has already been given in \cite{boost}, where a toy
model for these objects was constructed. The large radius limit of
these higher dimensional black rings will naturally be a boosted
Myers-Perry black string. We have searched for a similar
instability in these geometries but we have found no superradiant
instability. The stability of the Myers-Perry black string against this
mechanism was established by Cardoso and Yoshida
\cite{cardosobraneinst2}. We have concluded that boosting
the Myers-Perry black string does not change its superradiant
stability.
Again, the best way to understand the reason is to
look to the Schr$\ddot{\rm o}$dinger-like wave equation and its
factorized potentials as was done in section  \ref{sec:Hi dim}.
Although superradiance is still possible, the instability for
$D>5$ is not present because the KK momentum is not able to
produce trapped bound states. This is clearly seen in the
Schr$\ddot{\rm o}$dinger factorized potentials shown in Fig.
\ref{fig:potential HiD}.(a). Again we can extend the analysis to
black rings with finite radius. A similar reasoning as the one of
the previous paragraph together with the absence of instability
leads to the prediction that higher dimensional doubly (multiply)
spinning black rings will not have a superradiant upper bound for
their rotations along  $S^{n+2}$. The situation is somewhat
similar to the Kerr/Myers-Perry black hole: the Kerr and 5D
Myers-Perry solutions have an upper rotation bound but $D\geq 6$
Myers-Perry black holes do not.  However, we expect that another
dynamical process -- the ultra-spin mechanism \cite{ultra} --
might introduce a bound in the rotation.

It is also being conjectured that more general higher dimensional
black rings with other topologies rather than $S^1 \times S^{n+2}$
might exist \cite{ElvObers}. An example would be a black object with topology
$S^{\frac{D-3}{2}}\times S^{\frac{D-1}{2}}$ in odd $D$ dimensions.
In principle the large radius limit of these multiply spinning
black objects will be a boosted Myers-Perry membrane, i.e., (in
the previous example) a $\left ( \frac{D+3}{2}
\right)$-dimensional Myers-Perry black hole extended along
$\frac{D-3}{2}$ flat boosted directions. These black rings will
also be stable against superradiant bound modes since the KK
contribution, due to the $\left( \frac{D-3}{2} \right)$-brane on
the wave equation, is not effectively different from the one
coming from a single line. Indeed, the only difference in the
potential (\ref{potential U hiDIM}) is that the KK massive term is
$\kappa^2=\sum \kappa_i^2$ with $i=1,\cdots,(D-3)/2$, but this
does not change the stability results \cite{cardosobraneinst2}.

\section*{Acknowledgements}

It is a pleasure to thank the participants of the KITP Program:
``Scanning New Horizons: GR Beyond 4 Dimensions", and in
particular Vitor Cardoso, Roberto Emparan, Donald Marolf, Robert
Myers and Simon Ross for discussions and comments. Research at the
Perimeter Institute is supported in part by funds from NSERC of
Canada and MEDT of Ontario. I acknowledge financial support from
Funda\c c\~ao para a Ci\^encia e Tecnologia (FCT) - Portugal
through grant SFRH/BPD/2004, and the support of a NSERC Discovery
grant through the University of Waterloo. I would also like to
thank KITP and CENTRA for hospitality. This research was supported
in part by the National Science Foundation under Grant No.
PHY99-07949.

\appendix
\section{\label{sec:A1}Gamma function relations}

In this appendix we derive some gamma functions relations that are
used in the main body of the text.

i) We start with the relations needed to extract $\delta \rho$ from
 the gamma functions that appear in (\ref{far field-small x aux})
 in order to get (\ref{far field-small x}).

The properties $M(\tilde{a},\tilde{b},\chi=0)=1$ and
$\Gamma(x)\Gamma(1-x)=\pi /\sin(\pi x)$ (with
$x=\tilde{b}-\tilde{a}$ and then with $x=\tilde{b}-1$) allow us to
write (\ref{far field-small x aux}) as
\begin{eqnarray}
\Psi \!\!&\sim&\!\! \beta \, \chi^{l} e^{-\chi/2}\, {\biggl [}
\frac{\sin[\pi (\tilde{b}-\tilde{a})]}{\sin(\pi \tilde{b})}
\frac{\Gamma(\tilde{b}-\tilde{a})}{\Gamma(\tilde{b})}
+\chi^{1-\tilde{b}}\,
\frac{\Gamma(\tilde{b}-1)}{\Gamma(\tilde{a})} {\biggr ]}.\nonumber \\
& &
 \label{far small x A1}
\end{eqnarray}
Use of (\ref{Wittaker parameters end}) with $\delta\rho \sim 0$
yields
\begin{eqnarray}
\frac{\sin[\pi (\tilde{b}-\tilde{a})]}{\sin(\pi \tilde{b})}
\frac{\Gamma(\tilde{b}-\tilde{a})}{\Gamma(\tilde{b})} \sim (-1)^n
\frac{(2l+1+n)!}{(2l+1)!}\,.
 \label{far small x A2}
\end{eqnarray}
To simplify the second term in between brackets in
 (\ref{far small x A1}), we use $\Gamma(\tilde{a})\Gamma(1-\tilde{a})=\pi /\sin(\pi \tilde{a})$
with $\tilde{a}$ defined in (\ref{Wittaker parameters end}) to get
\begin{eqnarray}
& & \frac{1}{\Gamma(-n-i\delta \rho)} \sim -\frac{n!}{\pi}
 \sin[\pi (n+i\delta \rho)] \sim (-1)^{n+1} n! i\delta \rho \,, \nonumber \\
& &
 \label{far small x A3}
\end{eqnarray}
where in the first approximation we used  $\delta \rho \sim 0$ to
obtain $\Gamma(1+n+i\delta \rho) \sim \Gamma(1+n)$, and in the
second approximation we used $\sin(x+y)=\sin x \cos y+ \cos x \sin
y$ together with $\sin(i \pi \delta\rho)\sim i \pi \delta\rho$
(valid for small $\delta \rho$).

The relations  (\ref{far small x A1})-(\ref{far small x A3}) allow
us to go from (\ref{far field-small x aux}) into
 (\ref{far field-small x}).

ii) Finally, we derive the relations needed to make the transition
from (\ref{Matching Instab})
 into (\ref{delta rho}). Use of
$\Gamma(1+x)=x\Gamma(x)$ yields
\begin{eqnarray}
  & & \frac{\Gamma(l+1- i \,2\varpi)}
   {\Gamma(-l- i  \,2\varpi)}
   = i \,(-1)^{l+1}2\varpi \prod_{\jmath=1}^l
(\jmath^2+4\varpi^2)\,, \nonumber \\
& & \frac{\Gamma(-2l-1)}{\Gamma(-l)}=(-1)^{l+1}
\frac{l!}{(2l+1)!}\,. \label{gamma values 3}
\end{eqnarray}

\section{\label{sec:A2}Analytical results for the unboosted case}

In sections \ref{sec:geometries}-\ref{sec:quantitative}, we have
studied the instability of a {\it massless} scalar field with KK
momentum in the the boosted Kerr black string geometry. When we
{\it turn off} the boost,  the problem is equivalent to the
instability of a {\it massive} scalar field in the Kerr black hole
background, as long as we identify the KK momentum with the mass
of the scalar field. This last problem was studied originally by
Detweiler \cite{detweiler}. To clarify this equivalence, and as a
check of our results, in this appendix we show that our analysis
yields the results of  \cite{detweiler} when we set $\sigma=0$. In
this case we can give the quantitative features of the instability
in an approximated analytical form.

When the boost is switched off,  the second relation of (\ref{far
wave parameters}) becomes simply,
\begin{eqnarray}
 \rho=\frac{M \kappa^2}{\sqrt{\kappa^2-\omega^2}}\,, \label{B1}
 \end{eqnarray}
Use of (\ref{frequency spectrum}) in (\ref{B1}), yields $\frac{M
\kappa^2}{\sqrt{\kappa^2-\omega^2}}=l+N+1+i\delta\rho$. Letting
$\omega=\omega_{\rm R}+i\omega_{\rm I}$, one has
$\kappa^2-\omega^2 \sim \kappa^2-\omega_{\rm R}^2$ since
$\omega_{\rm I} \ll \omega_{\rm R}$. Moreover, taking $\delta\rho
\ll l+N+1$ one gets
 \begin{eqnarray}
\omega_{\rm R}^2\approx \kappa^2 \left[ 1-\left(\frac{\kappa
M}{l+n+1}\right)^2\right]\,, \label{B2}
 \end{eqnarray}
and thus  $\omega_{\rm R}\sim \kappa$ since the results are valid
in the regime $\kappa M \ll 1$. Moreover, differentiating
(\ref{B1}), $\delta \rho \approx M\kappa^2
(\kappa^2-\omega^2)^{-3/2}\omega \delta \omega$, and use of
$\omega_{\rm R}\sim \kappa$, $\delta \omega \equiv \omega_{\rm I}$
and (\ref{B2}) yields
\begin{eqnarray}
\omega_{\rm I}\approx \frac{\delta\rho}{M} \left(\frac{\kappa
M}{l+n+1}\right)^3\,, \label{B3}
 \end{eqnarray}
with $\delta\rho$ given by (\ref{delta rho}). Expressions
(\ref{B2}) and (\ref{B3}) are exactly the relations for the real
and imaginary part of the unstable modes found originally by
Detweiler \cite{detweiler}, for a massive scalar field with mass
$\kappa$ propagating in the unboosted Kerr background.



\begin{thebibliography}{99}
\bibitem{typo} There are two small typos in
\cite{cardosobraneinst2} that do not change the physical results:
One is in the last line of (A.7), and Fig. 7.b does not have a
potential well. I thank V. Cardoso and S. Yoshida for pointing out
them.
\end{thebibliography}
\end{document}